\documentclass[fleqn,usenatbib,useAMS]{pasa}%

\usepackage{float}
\usepackage{graphicx}	
\usepackage{amsmath}	
\usepackage{amssymb}	
\usepackage{multicol}        
\usepackage{bm}		
\usepackage{pdflscape}	
\usepackage{multirow}
\usepackage{subcaption}
\usepackage{xcolor}

\newcommand{\myaffil}[1]{$^{\rm #1}$}
\newcounter{inst}
\newcommand{\inst}[1]{\noindent%
  \refstepcounter{inst}\myaffil{\arabic{inst}\label{#1}}     
   }

\title[Remnant Radio Galaxies Discovered in a Multi-frequency Survey]{Remnant Radio Galaxies Discovered in a Multi-frequency Survey}

\author[Quici et al.]{

B.~Quici\myaffil{\ref{ICRAR}},
N.~Hurley-Walker\myaffil{\ref{ICRAR}},
N.~Seymour\myaffil{\ref{ICRAR}},
R.~J.~Turner\myaffil{\ref{uTAS}},
S.~S.~Shabala\myaffil{\ref{uTAS},\ref{ASTRO3D}},
M.~Huynh\myaffil{\ref{CSIRO}},
H.~Andernach\myaffil{\ref{mexico}},
A.~D.~Kapi\'nska\myaffil{\ref{NRAO}},
J.~D.~Collier\myaffil{\ref{CASS},\ref{wSYD},\ref{ct}}
M.~Johnston-Hollitt\myaffil{\ref{ICRAR}},
S.~V.~White\myaffil{\ref{ICRAR},\ref{uSAF}},
I.~Prandoni\myaffil{\ref{inaf}},
T.~J.~Galvin\myaffil{\ref{CSIRO}},
T.~Franzen\myaffil{\ref{ASTRON},\ref{ICRAR}},
C.~H.~Ishwara-Chandra\myaffil{\ref{TIFR}},
S.~Bellstedt\myaffil{\ref{UWA}},
S.~J.~Tingay\myaffil{\ref{ICRAR}},
B.~M.~Gaensler\myaffil{\ref{dunlap}},
A.~O'Brien\myaffil{\ref{CSIRO_},\ref{CFG},\ref{wSYD}},
J.~Rogers\myaffil{\ref{uTAS}},
K.~Chow\myaffil{\ref{CSIRO_}},
S.~Driver\myaffil{\ref{UWA}}, and
A.~Robotham\myaffil{\ref{UWA}}

{\scriptsize\inst{ICRAR}\,International Centre for Radio Astronomy Research, Curtin University, Bentley, WA 6102, Australia}\\
{\scriptsize\inst{uTAS}\,School of Natural Sciences, University of Tasmania, Private Bag 37, Hobart, 7001, Australia}\\
{\scriptsize\inst{ASTRO3D}\,ARC  Centre  of  Excellence  for  All  Sky  Astrophysics  in  3  Dimensions  (ASTRO  3D)}\\
{\scriptsize\inst{CSIRO}\,CSIRO Astronomy and Space Science, 26 Dick Perry Avenue, Kensington, WA 6151, Australia}\\
{\scriptsize\inst{mexico}\,IFUG, Universidad de Guanajuato, Guanajuato, C.P. 36000, Mexico"  for "DCNE, Universidad de Guanajuato, Cj\'on.\ de Jalisco s/n, Guanajuato, C.P. 36023, Mexico}\\
{\scriptsize\inst{NRAO}\,National Radio Astronomy Observatory, 1003 Lopezville Road, Socorro, NM 87801, USA}\\
{\scriptsize\inst{CASS}\,CSIRO Astronomy and Space Science (CASS), Marsfield, NSW 2122, Australia}\\
{\scriptsize\inst{wSYD}\,School of Science, Western Sydney University, Locked Bag 1797, Penrith, NSW 2751, Australia}\\
{\scriptsize\inst{ct}\,The Inter-University Institute for Data Intensive Astronomy (IDIA), Department of Astronomy, University of Cape Town, Private Bag X3, Rondebosch, 7701, South Africa}
{\scriptsize\inst{uSAF}\,Department of Physics and Electronics, Rhodes University, PO Box 94, Grahamstown, 6140, South Africa}\\
{\scriptsize\inst{inaf}\,Istituto di Radioastronomia, Via P. Gobetti 101, 40129, Italy}\\
{\scriptsize\inst{ASTRON}\,ASTRON: the Netherlands Institute for Radio Astronomy, PO Box 2, 7990 AA, Dwingeloo, The Netherlands}\\
{\scriptsize\inst{TIFR}\,National Centre for Radio Astrophysics, TIFR, Post Bag No. 3, Ganeshkhind Post, 411007 Pune, India}\\
{\scriptsize\inst{UWA}\,International Centre for Radio Astronomy Research, M468, University of Western Australia, Crawley, WA 6009, Australia}\\
{\scriptsize\inst{dunlap}\,Dunlap Institute for Astronomy and Astrophysics, University of Toronto, 50 St. George Street, Toronto ON M5S 3H4, Canada}\\
{\scriptsize\inst{CSIRO_}\,CSIRO Astronomy and Space Science, PO Box 76, 1710, Epping, NSW, Australia}\\
{\scriptsize\inst{CFG}\,{Center for Gravitation, Cosmology, and Astrophysics, Department of Physics, University of Wisconsin-Milwaukee, P.O. Box 413, Milwaukee, WI 53201, USA}}
}

\jid{PASA}
\doi{10.1017/pas.\the\year.xxx}
\jyear{\the\year}

\usepackage{aas_macros}
\usepackage{hyperref} 
\hypersetup{colorlinks,citecolor=blue,linkcolor=blue,urlcolor=blue}

\begin{document}

\begin{frontmatter}
\maketitle

\begin{abstract}
The remnant phase of a radio galaxy begins when the jets launched from an active galactic nucleus are switched off. To study the fraction of radio galaxies in a remnant phase, we take advantage of a $8.31$\,deg$^2$ sub-region of the GAMA~23~field which comprises of surveys covering the frequency range 0.1--9\,GHz. We present a sample of 104 radio galaxies compiled from observations conducted by the Murchison Wide-field Array (216\,MHz), the Australia Square Kilometer Array Pathfinder (887\,MHz), and the Australia Telescope Compact Array (5.5\,GHz). We adopt an `absent radio core' criterion to identify 10 radio galaxies showing no evidence for an active nucleus. We classify these as new candidate remnant radio galaxies. Seven of these objects still display compact emitting regions within the lobes at 5.5\,GHz; at this frequency the emission is short-lived, implying a recent jet switch-off. On the other hand, only three show evidence of aged lobe plasma by the presence of an ultra-steep spectrum ($\alpha<-1.2$) and a diffuse, low surface-brightness radio morphology. The predominant fraction of young remnants is consistent with a rapid fading during the remnant phase. Within our sample of radio galaxies, our observations constrain the remnant fraction to $4\%\lesssim f_{\mathrm{rem}} \lesssim 10\%$; the lower limit comes from the limiting case in which all remnant candidates with hotspots are simply active radio galaxies with faint, undetected radio cores. Finally, we model the synchrotron spectrum arising from a hotspot to show they can persist for 5--10\,Myr at 5.5\,GHz after the jets switch off -- radio emission arising from such hotspots can therefore be expected in an appreciable fraction of genuine remnants.

\end{abstract}
\end{frontmatter}

\section{Introduction}
\label{sec:remevolution}
The jets launched from a radio-loud active galactic nucleus (AGN) arise from the accretion onto a super-massive black hole, and form synchrotron-emitting radio lobes in the intergalactic environments of their host galaxies \citep{1974MNRAS.166..513S}. Whilst the jets are active, a radio galaxy will often display compact features such as an unresolved radio core coincident with its host galaxy, bi-polar jets, and hotspots in the lobes of Fanaroff Riley type II \citep[ FR-II;][]{1974MNRAS.167P..31F} radio galaxies. The radio continuum spectrum arising from the lobes is usually well approximated by a broken power-law for radio frequencies between $100$\,MHz and $10$\,GHz; the observed spectral index, $\alpha$\footnote{The spectral index $\alpha$ is defined through $S_\nu \propto \nu^\alpha$}, typically ranges within $-1.0<\alpha<-0.5$. Steepening in the radio lobe spectrum is allowed by $\Delta \alpha\leq0.5$ \citep[e.g. the CI model;][]{1962SvA.....6..317K,1970ranp.book.....P,1973A&A....26..423J}, and is attributed to ageing of the lobe plasma.

Such active radio galaxies do not offer a complete picture to the life-cycle of radio galaxies, due to a seemingly intermittent behaviour of the AGN jet activity. The remnant phase of a radio galaxy begins once the jets switch off. During this phase the lobes will fade as they undergo a rapid spectral evolution, e.g. as shown observationally by \citet{2011A&A...526A.148M}, \citet{2012A&A...545A..91S}, \citet{2017MNRAS.471..891G}, \citet{2017A&A...606A..98B}, \citet{2018MNRAS.475.4557M}, \citet{2020arXiv200409118J}, and shown with modelling conducted by \citet{2018MNRAS.476.2522T}, \citet{2018MNRAS.475.2768H} and \citet{2020MNRAS.tmp.1303S}. Remnant radio galaxies, remnants herein, will remain observable for many tens of Myr at low ($\sim150$\,MHz) frequencies, which is comparable but shorter than the duration of their previous active phase \citep{2017A&A...600A..65S,2018MNRAS.476.2522T,2020arXiv200313476B}. Jets are also known to restart after a period of inactivity \citep[e.g;][]{1994ApJ...421L..23R}, giving rise to a restarted radio galaxy. Several observational classes exist to describe such sources; double-double radio galaxies \citep[DDRG;][]{2000MNRAS.315..371S} describe sources in which two distinct pairs of lobes can be observed, however restarting jets can also appear as compact steep spectrum sources embedded within larger-scale remnant lobes \citep[e.g;][]{2018A&A...618A..45B}.

Compiling samples of remnant \citep{2012ApJS..199...27S,2017MNRAS.471..891G,2017A&A...606A..98B,2018MNRAS.475.4557M} and restarted \citep{2012ApJS..199...27S,2019A&A...622A..13M} radio galaxies sheds new light on their dynamics and evolution, and by extension, the AGN jet duty cycle. \citet{2020arXiv200409118J} present a direct analysis of the radio galaxy life cycle, in which their sample is decomposed into active, remnant and restarted radio galaxies. To complement these observational works, \citet{2020MNRAS.tmp.1303S} present a new methodology in which uniformly-selected samples of active, remnant and restarted radio galaxies are used to constrain evolutionary models describing the AGN jet duty cycle.

However, using radio observations to confidently identify radio sources in these phases is a challenging task, even in the modern era of radio instruments. Remnant radio galaxies, which are the focus of this work, display various observational properties that correlate with age, presenting a challenge for identifying complete samples of such sources. Various selection techniques exist amongst the literature, each with their selection biases. Due to the red preferential cooling of higher-energy synchrotron-radiating electrons \citep{1973A&A....26..423J,1994A&A...285...27K}, many authors have identified remnants by their ultra-steep radio spectrum ($\alpha<-1.2$), reflecting the absence of a source of energy injection, e.g. \citep{1987MNRAS.227..695C,2007A&A...470..875P,2015MNRAS.447.2468H}. However, \citet{2016A&A...585A..29B} demonstrates that this technique preferentially selects aged remnants and will miss those in which the lobes have not had time to steepen over the observed frequency range. \citet{2011A&A...526A.148M} propose a spectral curvature (SPC) criterion, which evaluates the difference in spectral index over two frequency ranges, e.g. SPC~=~$\alpha_{\mathrm{high}}$ - $\alpha_{\mathrm{low}}$ such that SPC~$>$0 for a convex spectrum. Sources with SPC~$>0.5$ demonstrate highly curved spectra, and are likely attributed to remnant lobes. However, modelling conducted by \citet{2017MNRAS.471..891G} shows that not all remnants will be selected this way, even at high ($\sim$10 GHz) frequencies.

Morphological selection offers a complementary way to identify remnants, independent of spectral ageing of the lobes. This technique often involves searching for low surface brightness (SB) profiles (SB~$<50$\,mJy arcmin$^{-2}$), amorphous radio morphologies, and an absence of hotspots \citep[e.g;][]{2012ApJS..199...27S,2017A&A...606A..98B}. However, young remnants in which the hotspots have not yet disappeared due to a recent switch-off of the jets, \citep[e.g 3C~028;][]{1984A&A...139...50F,2015MNRAS.454.3403H}, will be missed with these techniques. 

An alternative approach is to identify remnants based on an absent radio core; a radio core should be absent if the AGN is currently inactive \citep{1988A&A...199...73G}. This property is often invoked to confirm the status of a remnant radio galaxy, e.g. see \citet{1987MNRAS.227..695C}, and is recently employed by \citet{2018MNRAS.475.4557M} as a criterion to search for remnant candidates in a LOw Frequency ARray (LOFAR)-selected sample. The caveat here is the plausibility for a faint radio core to exist below the sensitivity of the observations, meaning this method selects only for remnant candidates. This method will also miss remnant lobes from a previous epoch of activity in restarted radio galaxies. A likely example of such a source is MIDAS~J230304-323228, discussed in Sect~\ref{sec:methodology}. These sources are beyond the scope of this work, however are a promising avenue for future work. 

A common aspect of almost all previously-mentioned observational studies of remnant radio galaxies, is their selection at low frequencies (typically around 150\,MHz). The preferential radiating of high-energy electrons means that the observable lifetime of remnant lobes increases at a lower observing frequency. Incorporating a low-frequency selection thus plays an important role in improving the completeness of remnant radio galaxy samples; quite often the oldest remnant lobes are detectable only at such low frequencies.

In this work, we take advantage of a broad range of radio surveys targeting the Galaxy~And~Mass~Assembly \citep[GAMA;][]{2011MNRAS.413..971D}~23 field to identify and study new remnant radio galaxies candidates in which the central AGN is currently inactive. We use new low-frequency observations provided by the Murchison Wide-field Array \citep[MWA; ][]{2013PASA...30....7T} and the Australian Square Kilometre Array Pathfinder \citep[ASKAP;][]{2007PASA...24..174J,2016PASA...33...42M} to compile a sample of radio galaxies, and use high-frequency (5.5\,GHz) observations provided by the Australia Telescope Compact Array (ATCA) to identify remnant candidates by the `absent radio core' criterion. In Sect.~\ref{sec:data} we present the multi-wavelength data used for this work. In Sect.~\ref{sec:methodology} we discuss the compilation of our radio galaxy sample, the classification of remnant candidates, and the matching of host galaxies. In Sect.~\ref{sec:results} we discuss each of the selected remnant candidates. In Sect.~\ref{sec:discussion} we discuss the observed radio properties of the sample and emphasise the caveats of various remnant selection methods. We also discuss the rest frame properties of the sample, present the fraction of remnants constrained by our observations, and examine a particularly interesting remnant with detailed modelling. Finally, our results are concluded in Sect.~\ref{sec:conclusion}.

We assume a flat $\Lambda$CDM cosmology with $H_0=67.7$kms$^{-1}$Mpc$^{-1}$, $\Omega_{\rm M}$=0.307 and $\Omega_\Lambda$=1-$\Omega_{\rm M}$ \citep{2014A&A...571A..16P}. All coordinates are reported in the J2000 equinox.


\begin{figure}
    \centering
    \includegraphics[width=0.99\linewidth]{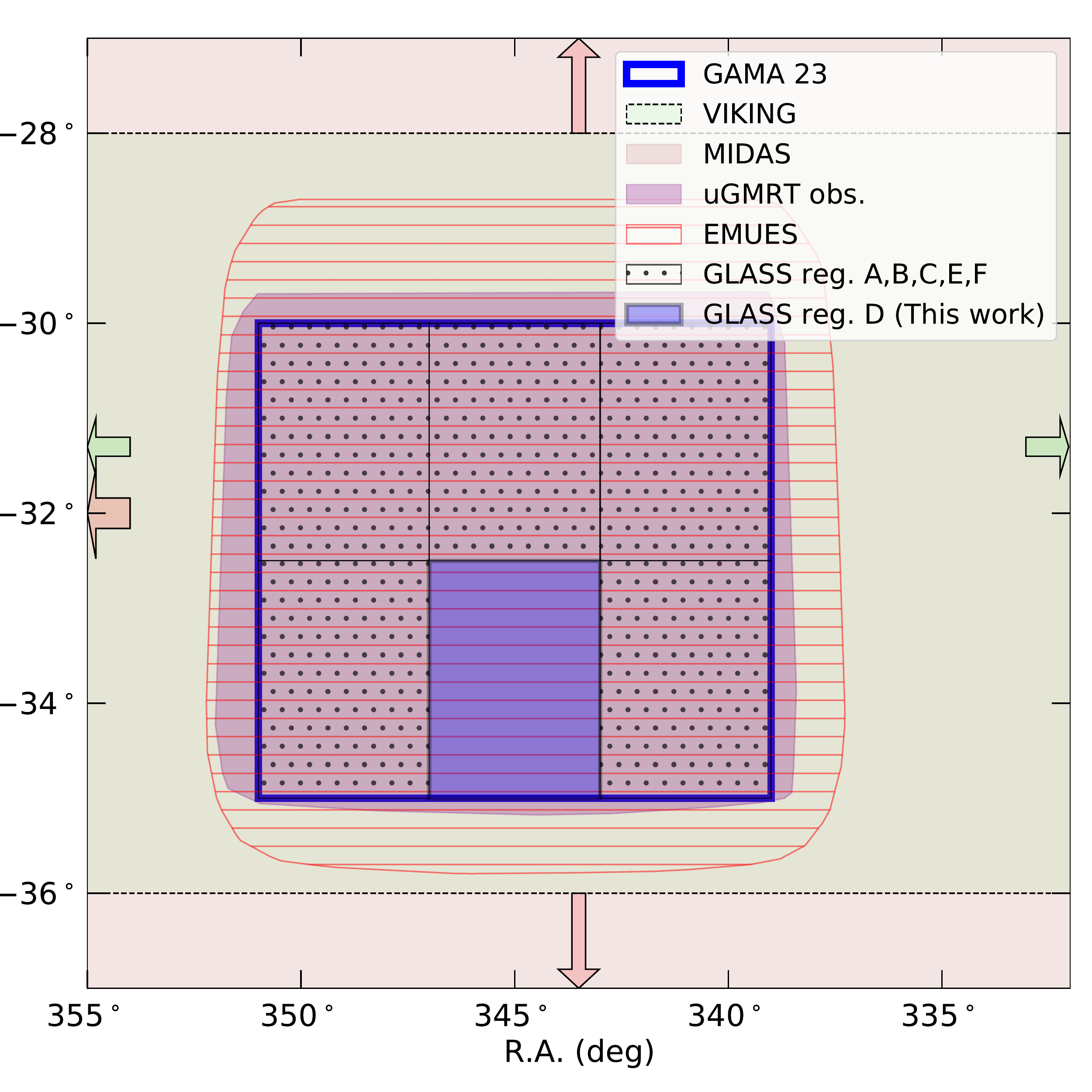}
    \caption{Sky coverage of radio surveys dedicated to observing GAMA~23, described in Sect.~\ref{sec:radiodata}. Near-infrared VIKING observations (Sect.~\ref{sec:vik}) are also displayed. Thick-red and thin-green arrows indicate the directions in which MIDAS~ES and VIKING extend beyond the represented footprint.}
    \label{fig:gama23cov}
\end{figure}
\section{DATA}
\label{sec:data}
Our sample is focused on a 8.31\,deg$^2$ sub-region of the GAMA~23 field ($\mathrm{RA}=345^\circ,\mathrm{Dec}=-32.5^\circ$), e.g. see Figure~\ref{fig:gama23cov}. The rich multi-wavelength coverage makes the absent radio-core criterion a viable method to search for remnant radio galaxies, and allows us to match the radio sample with their host galaxies.
\begin{table*}
\centering
\caption{Summarised properties of the radio surveys spanning the GAMA 23 field. Each column, in ascending order, details the telescope used to conduct the observations, the name of the radio survey, the dates observations were conducted, the central frequency of the observing band, the bandwidth available in each band, the average noise properties within region~D, and the shape of the restoring beam. }
\label{tab:radiodata}
\begin{tabular}{lccccccc}
\hline
Telescope & Survey & Date & Frequency & Bandwidth & Noise & Beam shape\\
& & observed &  \footnotesize{[MHz]}& \footnotesize{[MHz]} & \footnotesize{[mJy\,beam$^{-1}$]} & \footnotesize{Bmaj[$''$], Bmin[$''$], PA[$^\circ$}]\\
\hline
\multirow{3}{*}{\Bigg\{}\multirow{3}{*}{\footnotesize{MWA Phase I}}& \multirow{3}{*}{\footnotesize{GLEAM~SGP}} &\multirow{3}{*}{\footnotesize{2013--2015}} &  {\footnotesize{119}}  & \multirow{3}{*}{\footnotesize{30.72}} & {\footnotesize{16.2}}  & {\footnotesize{241, 202, -57}} \\
&  & & {\footnotesize{154}} & & \footnotesize{9.3} & \footnotesize{199, 157, -51}\\
&  & & {\footnotesize{186}} & & \footnotesize{6.7} & \footnotesize{162, 129, -47}\\
\footnotesize{MWA Phase II}& \footnotesize{MIDAS~ES} &\footnotesize{2018--2020} & \footnotesize{216} &\footnotesize{30.72}& \footnotesize{0.9} & \footnotesize{54, 43, 157} \\
\footnotesize{uGMRT} &\footnotesize{--} &\footnotesize{2016} & \footnotesize{399} &\footnotesize{200}&\footnotesize{0.1} & \footnotesize{15.8, 6.7, 9.5}\\
\footnotesize{ASKAP} & \footnotesize{EMU-ES} & \footnotesize{2019} & \footnotesize{887.5} & \footnotesize{288} & \footnotesize{0.035} & \footnotesize{10.5, 7.8, 87}\\
\footnotesize{VLA} & \footnotesize{NVSS} & \footnotesize{1993--1996} & \footnotesize{1400} &\footnotesize{50} & \footnotesize{0.45} & \footnotesize{45, 45, 0}\\
\footnotesize{ATCA} & \footnotesize{GLASS} & \footnotesize{2016--2020} & \footnotesize{5500, 9500} & \footnotesize{2000, 2000}& \footnotesize{0.024, 0.04} & \footnotesize{(4, 2, 0), (3.4, 1.7, 0)}\\
\hline
\end{tabular}
\end{table*}

\subsection{Radio data}
\label{sec:radiodata}
Below we describe the radio observations used throughout this work, summarised in Table~\ref{tab:radiodata} and~\ref{tab:atcadata}. 

\subsubsection{GLEAM SGP (119, 154, 185 MHz)}
\label{sec:GLEAM}
From 2013 to 2015, the MWA observed $\sim$30,000 deg$^2$ of the sky south of Dec~$= 30^\circ$. The GaLactic and Extra-galactic All-sky MWA \citep[GLEAM;][]{2015PASA...32...25W} survey adopted a drift scan observing method using the MWA Phase I configuration and a maximum baseline of $\sim$3\,km. Observations exclusive to 2013--2014 (year 1) were used to generate the public extra-galactic GLEAM first release \citep{2017MNRAS.464.1146H}\footnote{GLEAM catalogue publicly available on VizieR: \url{http://cdsarc.u-strasbg.fr/viz-bin/Cat?VIII/100}}. GLEAM spans a 72--231 MHz frequency range, divided evenly into 30.72 MHz wide-bands centered at $\nu_c=$ \big\{88, 119, 154, 185, 216\big\} MHz. 
Observations conducted during 2014--2015 (year 2) covered the same sky area as GLEAM, but with a factor of two increase in the observing time due to offset ($\pm$1 hour in median RA) scans. Within a footprint of: 310$^\circ \leq$ RA $\leq 76^\circ$, -48$^\circ \leq$ Dec $\leq -2^\circ$ ($\sim5,100$ deg$^2$ sky coverage), \textcolor{blue}{Franzen et al. (in prep)} independently reprocessed the overlapping observations from both years to produce the GLEAM South Galactic Pole (SGP) survey, achieving a factor $\sim3$ increase in integration time over GLEAM. Due to calibration errors associated with Fornax~A and Pictor~A entering the primary beam side-lobes, the lowest wide-band centered at 88\,MHz was discarded at imaging. Despite its declination-dependence, the GLEAM SGP root-mean-square (RMS) remains effectively constant within the footprint of our (8.31\,deg$^2$) study. The point spread function (PSF) varies slightly across the survey; at 216\,MHz the PSF size within GAMA~23 is approximately $150'' \times 119''$, and an RMS of $\sigma \sim 4\pm0.5$mJy beam$^{-1}$ is achieved. This is consistent with a factor$~2$ increase in sensitivity over GLEAM. We use an internal GLEAM~SGP component catalogue, developed by the method used for GLEAM. \textcolor{blue}{Franzen et al.}~(in prep) quote an 8\% absolute uncertainty in the GLEAM~SGP flux density scale, consistent with the value reported for GLEAM by \citet{2017MNRAS.464.1146H}.
\label{sec:dedicatedradio}

\subsubsection{MIDAS~ES (216 MHz)}
\label{sec:MIDAS}
In 2018 the MWA was reconfigured into an extended baseline configuration
\citep[MWA Phase II;][]{2018PASA...35...33W}.
The MWA Phase II provides a maximum baseline of 5.3\,km which, at 216\,MHz, improves the angular resolution to $54''\times43''$. This drives down the classical and sidelobe confusion limits, allowing for deeper imaging to be made with an RMS of $\sigma\ll 1\,$mJy\,beam$^{-1}$. MWA Phase II science is presented in \citet{2020PASA...37...14B}. The MWA Interestingly Deep Astrophysical Survey (MIDAS, Seymour et al. in prep) will provide deep observations of six extra-galactic survey fields, including GAMA~23, in the MWA Phase II configuration. 
Here, we use an early science (MIDAS~ES herein) image of
the highest frequency band centered at 216\,MHz.
Data reduction followed the  method outlined by \textcolor{blue}{Franzen et al. (in prep)} where each snapshot
was calibrated using a model
derived from GLEAM. 
Imaging was performed using the \textsc{WSClean} imaging software \citep{2014ascl.soft08023O} using a robustness of 0 \citep{1995AAS...18711202B}. Altogether, 54 two-minute snapshot images, each achieving an average RMS of $\sim 8\,$mJy\,beam$^{-1}$ without self calibration, were mosaiced
to produce a deep image. Stacking of individual snapshots reproduced the expected $\sqrt{t}$ increase in sensitivity, where $t$ is the total integration time, and implied the classical and sidelobe confusion limits were not reached. After a round of self calibration the final deep image, at zenith, achieved an RMS of $\sim$0.9\,mJy\,beam$^{-1}$ with a $\lesssim1'$ restoring beam. Radio sources were catalogued using the Background And Noise Estimation tool \textsc{bane}, to measure the direction-dependent noise across the image, and \textsc{aegean} to perform source finding and characterisation \citep{2012MNRAS.422.1812H,2018PASA...35...11H}\footnote{\textsc{aegean} and \textsc{bane} publicly available on GitHub: \url{https://github.com/PaulHancock/Aegean}. This work made use of 2.0.2 version of \textsc{aegean}}. Using the GLEAM catalogue, sources above 10$\sigma$ in MIDAS~ES and GLEAM were used to correct the MIDAS~ES flux-scale. For 887 such sources in GAMA~23, correction factors were derived based on the integrated flux density ratio between MIDAS~ES and GLEAM. The correction factors followed a Gaussian distribution, and indicated an agreement with GLEAM within 5\%. Given the 8\% uncertainty in the GLEAM flux density scale (Sect.~\ref{sec:GLEAM}), we prescribe an 8\% uncertainty in the MIDAS~ES flux density scale.

\subsubsection{EMU Early Science (887 MHz)}
In 2019 the ASKAP delivered the first observations conducted with the full 36-antenna array configuration. The design of ASKAP opens up a unique parameter space for studying the extra-galactic radio source population. With a maximum baseline of 6\,km, ASKAP produces images with $\sim 10''$ resolution at 887\,MHz. The shortest baseline is 22\,m, recovering maximum angular scales of $\sim0.8^\circ$. As part of the Evolutionary Map of the Universe \citep[EMU;][]{2011JApA...32..599N} Early Science, the GAMA~23 field was observed at 887\,MHz and made publicly available on the CSIRO ASKAP Science Data Archive (CASDA\footnote{\url{https://data.csiro.au/collections/domain/casdaObservation/search/}}). Details of the reduction are briefly summarized here. This data has been reduced by the ASKAP collaboration using the \textsc{ASKAPsoft}\footnote{\url{https://www.atnf.csiro.au/computing/software/askapsoft/sdp/docs/current/pipelines/introduction.html}} data-reduction pipeline (\textcolor{blue}{Whiting et al. in prep}) on the Galaxy supercomputer hosted by the Pawsey Supercomputing Centre. PKS~B1934-638 was used to perform bandpass and flux calibration for each of the unique 36 beams. Bandpass solutions were applied to the target fields. The final images are restored with a 10.55$'' \times 7.82''$  (BPA = 86.8$^\circ$) elliptical beam, and achieves an RMS of $\sigma \approx 34-45\mu$Jy beam$^{-1}$. Henceforth we refer to these observations as EMU-ES.
\begin{table*}
\centering
\caption{Details of the additional ATCA data collected here under project code C3335, PI: B. Quici. Each column, in ascending order, details the configuration used to conduct observations, the date observations were conducted, the central frequency of the receiver band, the bandwidth available within each band, the approximate time spent per-source in each configuration, the secondary calibrator observed, the average noise per pointing, and the shape of the restoring beam. Note, primary calibrator PKS~B1934-638 was observed for all observations. }
\label{tab:atcadata}
\begin{tabular}{lcccccccc}
\hline
Config. & Date & Frequency& Bandwidth & Duration  & Secondary & Noise & Beam shape \\
& observed& [GHz]& [GHz] &[min]& calibrator & [mJy beam$^{-1}$]&\footnotesize{Bmaj[$''$], Bmin[$''$], PA[$^\circ$}]\\
\hline
H168 & 18/10/19 & 2.1& 2 &22 & \footnotesize{PKS B2259-375} & 0.7 & 104, 127, 0\\
H75 & 24/10/19 & 5.5, 9.0& 2, 2 &25 &\footnotesize{PKS B2254-367} & 0.22, 0.14 & (88, 110, 0), (54, 67, 0) \\
\hline
\end{tabular}
\end{table*}
\subsubsection{NVSS (1.4 GHz)}
We use observations conducted by the National Radio Astronomy Observatory (NRAO) using the Very Large Array \citep[VLA;][]{1980ApJS...44..151T} to sample an intermediate frequency range. The NRAO VLA Sky Survey \citep[NVSS; ][]{1998AJ....115.1693C}\footnote{NVSS catalogue publicly available on VizieR: \url{https://vizier.u-strasbg.fr/viz-bin/VizieR?-source=\%20NVSS}} surveys the entire sky down to Dec$=-40^\circ$ at 1400\,MHz. Observations for NVSS were collected predominately in the D configuration, however the DnC configuration was used for Southern declinations. Final image products were restored using a circular synthesized beam of $45''\times45''$. At 1400\,MHz, NVSS achieves an RMS of $\sigma=0.45\,$mJy\,beam$^{-1}$.
\subsubsection{GLASS (5.5 \& 9.5 GHz)}
\label{sec:glass}
The GAMA Legacy ATCA Sky Survey (GLASS;~Huynh~et~al.~in prep) offers simultaneous, high-frequency (5.5,~9.5\,GHz) observations of GAMA 23 observed by the ATCA. Observations for GLASS were conducted over seven semesters between 2016 -- 2020 (PI: M. Huynh, project code: C3132). Data at each frequency were acquired with a 2\,GHz bandwidth, made possible by the Compact Array Broadband Backend \citep{2011MNRAS.416..832W}, with the correlator set to a 1\,MHz spectral resolution. Observations for GLASS were conducted in two separate ATCA array configurations, the 6A and 1.5C configurations, contributing 69\% and 31\% towards the total awarded time, respectively. The shortest interferometer spacing is 77\,m in the 1.5C configuration, providing a largest recoverable angular scale of 146$''$ at 5.5\,GHz. As part of the observing strategy, GLASS was divided into six 8.31\,deg$^2$ regions (regions~A~$-$~F), for which region~D (RA=345$^\circ$, Dec=-33.75$^\circ$) was observed, reduced and imaged first. For this reason, this paper focuses only on region~D.

Processing and data reduction were conducted using the Multichannel Image Reconstruction, Image Analysis and Display (\textsc{MIRIAD}) software package \citep{1995ASPC...77..433S}, similar to the method outlined by \cite{2015MNRAS.454..952H,2020MNRAS.491.3395H}. The 1435 region~D pointings are restored with the Gaussian fit of the dirty beam, and convolved to a common beam of $4''\times2''$ (BPA=$0^\circ$) at 5.5\,GHz, and achieves an RMS of $\sim 24\,\mu$Jy\,beam$^{-1}$. A similar process at 9.5\,GHz results in a $3.4''\times1.7''$ (BPA=$0^\circ$) synthesized beam and achieves an RMS of $\sim 40\,\mu$Jy\,beam$^{-1}$. Although the same theoretical sensitivity is expected at 5.5\,GHz and 9.5\,GHz, the sparse overlap in adjacent pointings, larger phase calibration errors, and increased radio frequency interference (RFI) all result in a drop in sensitivity. Henceforth, we refer to GLASS observations conducted at 5.5\,GHz and 9.5\,GHz as GLASS 5.5 and GLASS 9.5 respectively.
\subsubsection{uGMRT legacy observations (399 MHz)}
As part of the GLASS legacy survey (Sect.~\ref{sec:glass}), uGMRT has observed GAMA 23 in band-3 (250--500\,MHz) centered at 399\,MHz. The project $32\_060$ (PI: Ishwara Chandra) was awarded 33 hours to cover 50 contiguous pointings spanning a 50 square-degree region. Observations were conducted in a semi-snapshot mode, with $\approx 30$\,minutes per pointing distributed through three 10\,minute scans. In band-3, the wide-band correlator collects a bandwidth of 200\,MHz divided into 4,000 fine channels. Data reduction was conducted using a Common Astronomical Software Application \citep[CASA;][]{2007ASPC..376..127M} pipeline\footnote{Pipeline can be found at \url{http://www.ncra.tifr.res.in/~ishwar/pipeline.html}, and makes use of CASA version 5.1.2-4}. Data reduction followed the standard data reduction practices such as data flagging, bandpass and gain calibration, application of solutions to target scans, imaging and self-calibration (see \citealt{2020MNRAS.497.5383I} for details). The image is restored by a 15.8$''\times 6.71''$ (BPA= 9.5$^\circ$) beam, and achieves a best RMS of $\sim$100\,$\mu$\,Jy\,beam$^{-1}$. Note that several bright sources throughout the field adversely impact the data reduction, resulting in large spatial variations in the RMS. 
\subsubsection{Low resolution ATCA observations (2.1, 5.5 \& 9 GHz)}
\label{sec:LSCX} 
Due to their power-law spectral energy distributions, the integrated luminosities of radio galaxy lobes decrease with increasing frequency (Sect.~\ref{sec:remevolution}). Given that remnants can display ultra-steep radio spectra, this presents a sensitivity challenge associated with their detection. For a survey such as GLASS which has high $\theta\sim5''$ resolution (Sect.~\ref{sec:glass}), resolution bias further exacerbates this problem: diffuse low-surface-brightness regions escape detection with greater ease, resulting in an under-estimation of the integrated flux density. To combat the resolution bias suffered by GLASS, we carried out low-resolution observations with ATCA. 

The project C3335 (PI: B. Quici) was awarded 14 hours to conduct low-resolution observations of each remnant identified for this work, at 2.1, 5.5 and 9\,GHz. Observations at 2.1\,GHz (\textit{LS}-band) were conducted in the H168 configuration; observations at 5.5 and 9 GHz (\textit{CX}-band) were conducted in the H75 configuration. The minimum and maximum baselines achieved within each configuration are 61\,m and 192\,m for H168, 31\,m and 89\,m for H75 respectively (excluding baselines formed with antenna CA06). Bandpass, gain and flux calibration was performed using PKS~B1934-638. Due to the small angular separation between each target, the secondary calibrator was held constant for each target. At 2.1\,GHz, PKS~B2259-375 was used for phase calibration. At 5.5 and 9\,GHz, phase calibration was performed with PKS~B2254-367. To maximize $uv$ coverage, each target was observed on a rotating block of length $\sim$30\,minutes, with approximately\footnote{Time per source was varied slightly to accommodate for the fainter/brighter sources.} 2 minutes allocated per target per block. Secondary calibrators were observed twice per block to ensure stable phase solutions.

\begin{figure}
    \centering
    \includegraphics[width=\linewidth]{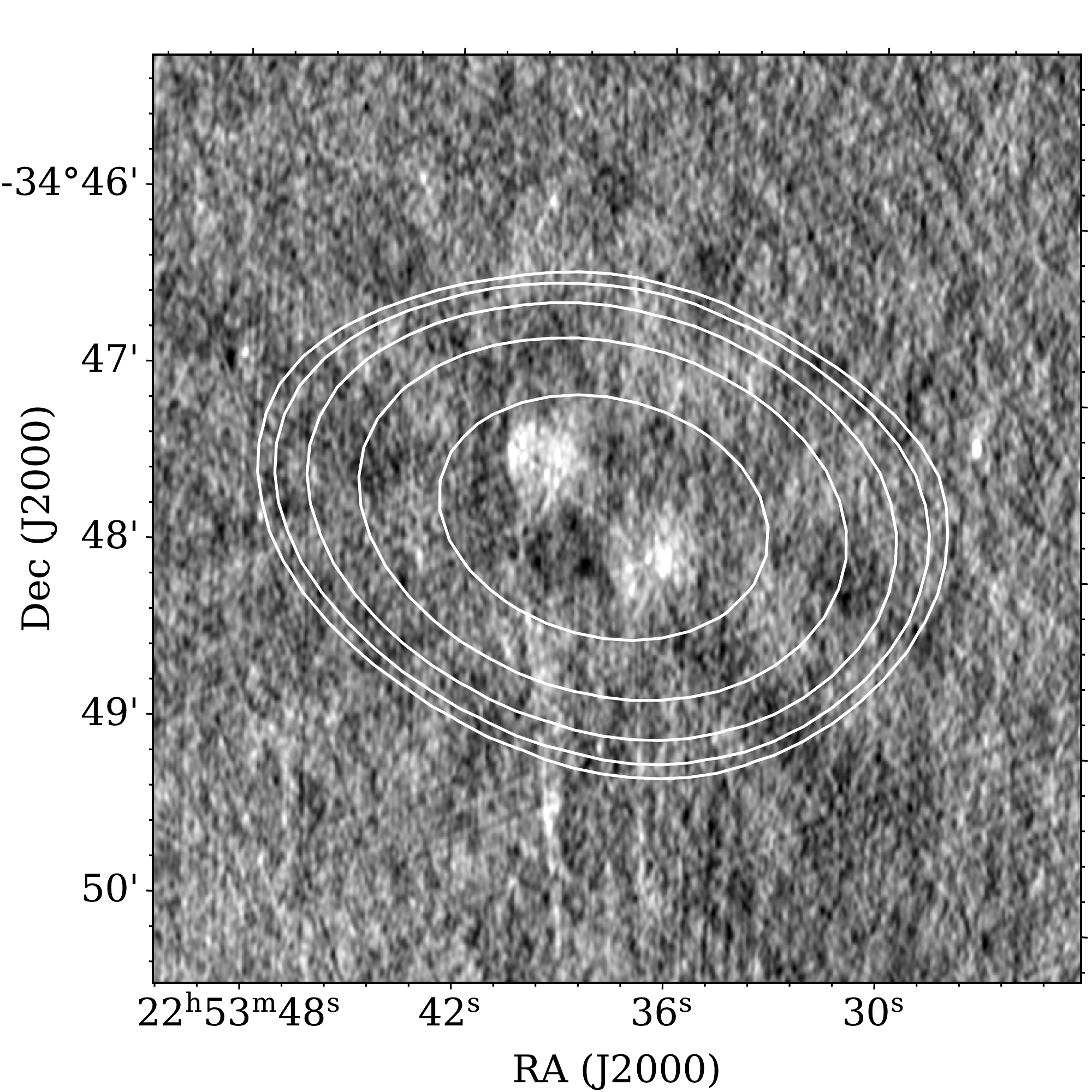}
    \caption{A 5.5\,GHz view of the radio source MIDAS~J225337$-$344745. The image offers a comparison between (i) GLASS (Sect.~\ref{sec:glass}) presented as the gray-scale image on a linear stretch, and (ii) the low-resolution ATCA observations (Sect.~\ref{sec:LSCX}) represented by the solid white contours. Contour levels are set at [5, 6.3, 9.6, 17.3, 35.5]$\times \sigma$, where $\sigma=210\mu$\,Jy\,beam$^{-1}$} and is the local RMS.
    \label{fig:glass_lowres_comparison}
\end{figure}

Data reduction was performed using the \textsc{CASA} software package\footnote{This work makes use of CASA version 5.1.2-4}, and followed the standard data reduction practices. As part of preliminary flagging, the data are `clipped', `shadowed' and `quacked' using the \texttt{flagdata} task, in order to flag for zero values, shadowed antennas and the first five seconds of the scans, respectively. Forty edge channels of the original 2049 are also flagged due to bandpass roll-off \citep{2011MNRAS.416..832W}. Again using the \texttt{flagdata} task, the uncalibrated data are then automatically flagged for RFI using \texttt{mode=`tfcrop'}, which calculates flags based on a time and frequency window. The data are manually inspected in an amplitude versus channel space, to ensure RFI is adequately flagged.  Observations conducted in the $LS$-band and $CX$-band were split into four and eight sub-bands, respectively. Calibration was performed per sub-band per pointing. To perform calibration, the complex gains and bandpass are solved for first using the primary calibrator. Complex gains and leakages were solved for next using the secondary calibrator. After applying a flux-scale correction based on PKS~B1934-638, the calibration solutions were copied individually to each target scan. A secondary round of automatic RFI flagging is performed with \texttt{flagdata} where \texttt{mode=`rflag'}, which is used for calibrated data. The primary and secondary calibrator, as well as each target scan are flagged in this way. In total, approximately $\approx52\%$ and $\approx 15\%$ of the available bandwidth was flagged due to RFI present in the $LS$ and $CX$ bands, respectively. Within the hybrid configurations the first five antennas, CA01--CA05, provide the dense packing of short (10--100\,m) spacings. Antenna CA06 is fixed and provides much larger spacings of $\sim$4500 m. Given that this results in a large gap in the $uv$ coverage, all baselines formed with antenna CA06 are excluded to achieve a well-behaved point-spread-function. The average noise properties and restoring beams at 2.1, 5.5 and 9\,GHz are respectively: $\sigma \sim0.7$\,mJy\,beam$^{-1}$ and $\theta=104''\times127''$,  $\sigma \sim0.22$\,mJy\,beam$^{-1}$ and $88''\times110''$, and $\sigma \sim0.14$\,mJy\,beam$^{-1}$ and $\theta=54''\times67''$. We use any unresolved GLASS sources present within the target scans to evaluate a calibration uncertainty. At 5.5\,GHz, the ratio of the integrated flux density between the low-resolution ATCA observations and GLASS was consistently within 3\%. We use this value as the absolute flux-scale uncertainty. Details of these observations are summarized in Table \ref{tab:atcadata}. A comparison of these observations and GLASS, at 5.5\,GHz, is presented in Figure~\ref{fig:glass_lowres_comparison}. The reader should note, due to persistent RFI in the $LS$-band, we were unable to make a detection of most of our targets at this frequency.
\subsection{Optical/near-infrared data}
\label{sec:nonradiodata}

\subsubsection{VIKING near-infrared imaging}
\label{sec:vik}
Observed with the Visible and Infrared Survey Telescope for Astronomy (VISTA), the VISTA Kilo-degree INfrared Galaxy survey (VIKING) provides medium-deep observations over $\sim$1500\,deg$^2$ across the $Z$, $Y$, $J$, $H$ \& $K_s$ bands, each achieving a 5$\sigma$ AB magnitude limit of 23.1, 22.3, 22.1, 21.5 and 21.2 respectively. VIKING imaging is used to perform both an automated and manual host galaxy identification (see Sect.~\ref{sec:methodology}). 

\subsubsection{GAMA 23 photometry catalogue}
The GAMA 23 photometry catalogue \citep{2020MNRAS.496.3235B} contains measured properties and classifications of each object catalogued by the \textsc{ProFound}\footnote{Publicly available on GitHub: \url{https://github.com/asgr/ProFound}} \citep{2018MNRAS.476.3137R} source finding routine, which uses a stacked $r+Z$ image to perform initial source finding. Approximately 48,000 objects have spectroscopic redshifts provided by the Anglo-Australian Telescope (AAT), and the survey is $\sim$95\% spectroscopically complete up to an $i$ band magnitude of 19.5 \citep{2015MNRAS.452.2087L}. For objects without spectroscopic redshifts, we use the near-UV to far-IR photometry (available in the photometry catalogue) to obtain photometric redshifts using a public photometric redshift code \citep[EAZY;][]{2008ApJ...686.1503B}\footnote{EAZY publicly available on GitHub: \url{https://github.com/gbrammer/eazy-photoz}}. 
\section{Methodology}
\label{sec:methodology}
\subsection{Sample construction}
\begin{table*}
\centering
\caption{Summary of each sample criteria discussed in Sect.~\ref{sec:methodology}. Steps 1--4 describe the radio galaxy sample selection. Step~5 describes the active/remnant classification. Step~6 describes host galaxy association.\\
$^\dagger$Step~3 is broken into two parts, denoted by steps~3.1~and~3.2, for which the resulting sample size is the sum total.}
\label{tab:methodology}
\begin{tabular}{lccc}
\hline

\hline
No.&Sample step & Criteria &  Sample size\\
\hline
\multirow{2}{*}{1}&\multirow{2}{*}{Limit footprint within GLASS region D} & 343$^\circ$ $\leq$ RA $\leq$ 347$^\circ$ & \multirow{2}{*}{676}\\
& & -32.5$^\circ$ $\leq$ Dec $\leq$ -35$^\circ$ & \\
2&Flux-density cut & $S_\mathrm{216MHz}>10\,$mJy & 446\\
3&Angular size cut$^\dagger$ & $\theta\geq25''$ & 109\\
\footnotesize{\textit{~~3.1}}&  &\footnotesize{$\theta_{\mathrm{GLASS}}\geq25''$} & \footnotesize{\textit{~~~~(82)}} \\
\footnotesize{\textit{~~3.2}}&  & \footnotesize{$\theta_{\mathrm{GLASS}}<25''$} \& $\theta_{\mathrm{EMU-ES}}\geq25''$& \footnotesize{{\textit{~~~~(27)}}}\\
\multirow{2}{*}{4}&\multirow{2}{*}{AGN dominated} & Remove radio sources tracing the  & \multirow{2}{*}{106}\\
&&optical/near-IR galaxy component &\\
\hline
\multirow{2}{*}{5}&\multirow{2}{*}{Activity status} & Radio core in GLASS~5.5 (Active) & 94 \\
& & No radio-core in GLASS~5.5 (Remnant cand.) & 10\\
\hline
\multirow{6}{*}{6}&\multirow{3}{*}{Host identification (Active)} &\multirow{3}{*}{Visually match radio core with VIKING galaxy}  & 26 \textit{($z_s$)}\\
& & & 54 \textit{($z_p$)}\\
& & & 14 \textit{(}no \textit{$z$)}\\\cline{2-4}
& \multirow{2}{*}{Host identification (Remnant)}& \multirow{2}{*}{See Sect.~\ref{sec:results}} & 3 \textit{($z_s$)}\\
& & & 7 \textit{($z_p$)} \\

\hline

\hline
\end{tabular}
\end{table*}
Our methodology for compiling a sample of radio galaxies, and classifying their activity status, is presented below. We begin our selection at low frequencies where steep-spectrum radio lobes are naturally brighter. Not only does MIDAS~ES provide access to such low frequencies, but also its relatively low spatial resolution results in a sensitivity to low-surface-brightness emission that is required to recover emission from diffuse, extended radio sources. 

Genuine remnant radio galaxies will not display radio emission from the core associated with the AGN jet activity at the centre of the host galaxy. As such, we classify any radio galaxy as `active' if positive radio emission is observed from the radio core. Similarly, we prescribe a `candidate remnant' status to any radio galaxy that demonstrates an absence of radio emission from the core. True emission from the radio core will be unresolved even on parsec scales, meaning observations with high spatial resolution are ideal for their detection. This also ensures that the emission from the radio core will not experience blending by the backflow of the radio lobes. As demonstrated by \citet{2018MNRAS.475.4557M}, sensitive observations are equally important to enable the detection of radio cores. GLASS addresses both of these considerations by providing sensitive ($\sim30\,\mu$Jy\,beam$^{-1}$), high resolution ($4''\times2''$) radio observations at 5.5\,GHz. Our sample is constructed by following the steps outlined below (see Table~\ref{tab:methodology}), and is presented in full as a supplementary electronic table (see Table~\ref{tab:supplementarytable} for column descriptions).

\begin{figure}
    \centering
    \includegraphics[width=\linewidth]{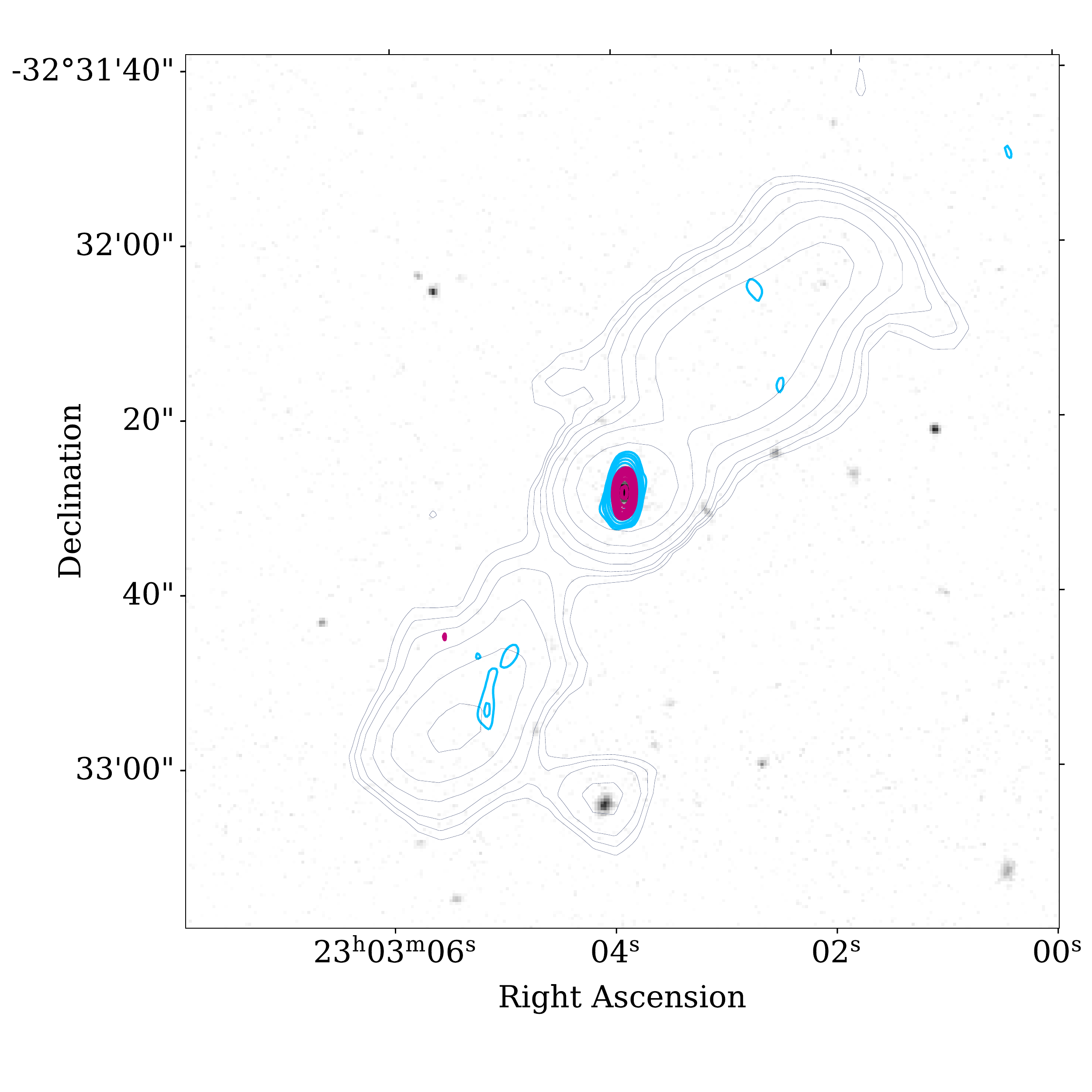}
    \caption{Example of the radio source MIDAS~J230304-323228 satisfying the criterion: $\theta_{\mathrm{GLASS}}<25''$ \& $\theta_{\mathrm{EMU-ES}}\geq25''$. The low-surface-brightness lobes are escaping detection in GLASS, resulting in an incomplete morphology. The contours represent EMU-ES (navy blue), GLASS~5.5 (cyan) and GLASS~9.5 (magenta), with levels set at [3,4,5,7,10,15,25,100]$\times \sigma$, where $\sigma$ is the local RMS of 43, 26 and 40\,$\mu$Jy\,beam$^{-1}$ respectively. Contours are overlaid on a linear stretch VIKING K$_s$-band image. The seemingly absent hotspots would imply these are remnant lobes, however the presence of a radio core means this source is classified as `active'. The true nature of this source \textit{may} be a restarted radio galaxy, however the lack of any resolved structure around the core is puzzling.}
    \label{fig:MIDAS_rest_example}
\end{figure}

\begin{figure}
    \centering
    \includegraphics[width=\linewidth]{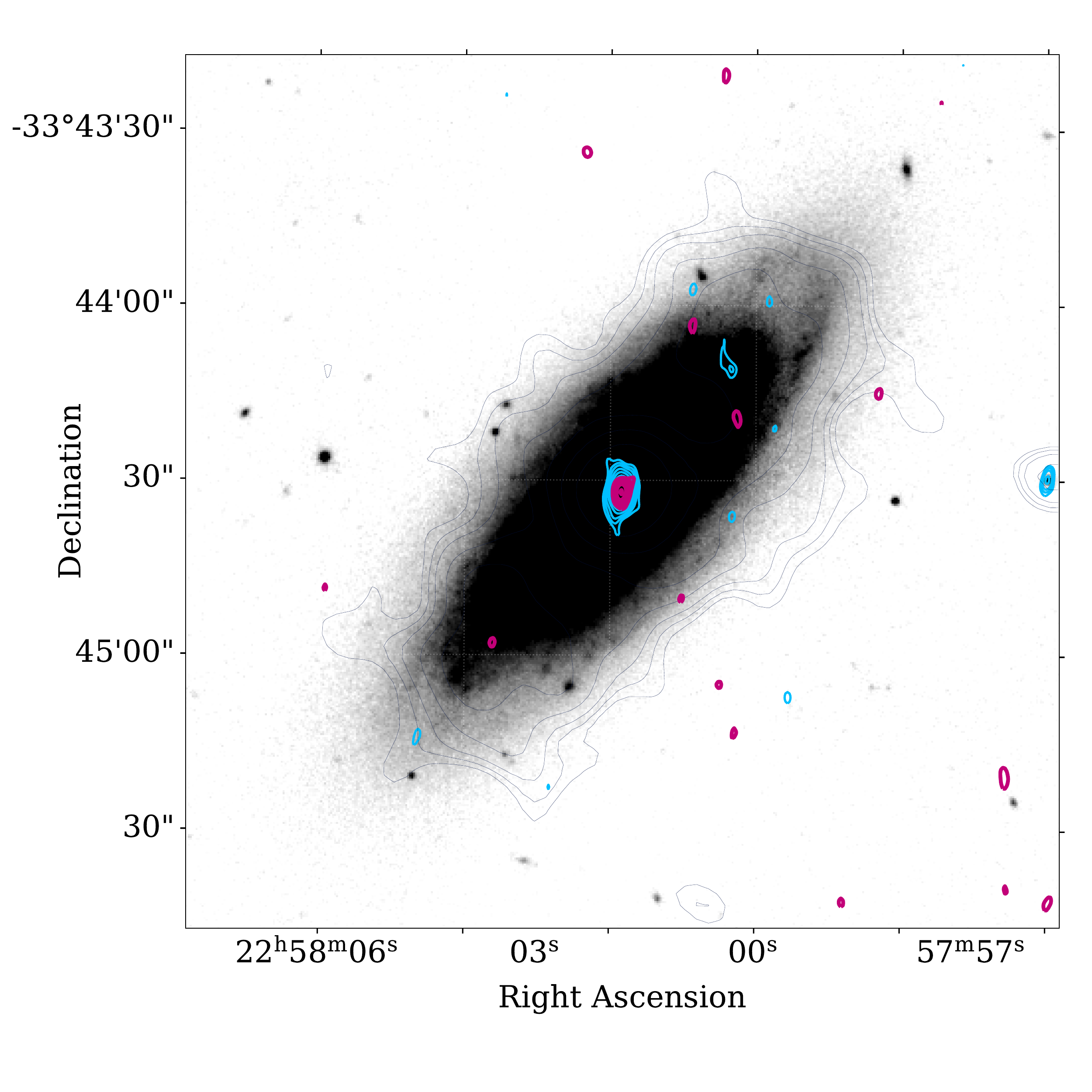}
    \caption{Example of a non AGN-dominated radio source, MIDAS~J225802-334432, excluded from the sample. Analysis of the radio morphology shows that the radio emission traces the optical component of the host galaxy. The contours represent EMU-ES (navy blue), GLASS~5.5 (cyan) and GLASS~9.5 (magenta), with levels set at [3,4,5,7,10,15,25,100]$\times \sigma$, where $\sigma$ is the local RMS of 45, 28 and 41\,$\mu$\,Jy\,beam$^{-1}$ respectively. Contours are overlaid on a linear stretch VIKING K$_s$-band image. The radio emission is hosted by IC 5271 (ESO 406-G34).}
    \label{fig:MIDAS_SFG}
\end{figure}

\begin{enumerate}
    \item \textit{Limit the search footprint within GLASS region D}\\
The footprint, within which radio galaxies are selected, is constrained to $343^\circ\leq \mathrm{RA} \leq~347^\circ$, $-35^\circ\leq \mathrm{Dec}\leq-32.5^\circ$. This excludes the outer regions of higher noise from the GLASS mosaic thus ensuring the GLASS noise levels range between 25$-$35$\mu$Jy~beam$^{-1}$ at 5.5\,GHz. By maintaining almost uniform noise levels across the field, this reduces the bias of selecting brighter radio cores in higher-noise regions. By applying this footprint to the MIDAS~ES component catalogue, 676 components are selected. The resulting sky coverage within this footprint is 8.31\,deg$^2$. 

    \item \textit{Flux-density cut}\\
A 10\,mJy flux density cut at 216\,MHz is imposed on the sample. Given the low angular resolution, this ensures the MIDAS~ES detections are robust, e.g. greater than 10$\sigma$ for an unresolved source. We identify 446 radio sources brighter than 10\,mJy at 216\,MHz.

    \item \textit{Angular size cut}\\
Our decision to impose a minimum size constraint was motivated by two factors: firstly to minimise blending of the radio core with the radio lobes, and secondly to allow for an interpretation of the radio source morphology. By imposing a minimum 25$''$ angular size constraint, we ensured a minimum of six GLASS~5.5 synthesized beams spread out across the source. For each of the 446 radio sources, we produce $2'\times 2'$ cutouts\footnote{Image cutouts are generated using \textsc{Astropy} module \texttt{Cutout2D}} centered at the catalogued MIDAS~ES source position. EMU-ES, GLASS~5.5, GLASS~9.5 and VIKING $K_s-$band cutouts are generated, and contours of the radio emission are overlaid onto the VIKING $K_s-$band image. Henceforth, we refer to these as image overlays. While GLASS~5.5 has the advantage in spatial resolution, EMU-ES has a significantly better brightness-temperature sensitivity. This consideration is important since faint radio lobes, while seen in EMU-ES, may become undetected by GLASS~5.5. 

As such, a first pass is conducted by identifying any radio source with an angular size greater than 25$''$ in GLASS~5.5 (e.g. $\theta_{\mathrm{GLASS~5.5}}\geq25''$). Due to the manageable size of the sample, we do this step manually by visually matching the correct components of each radio source, and measure the linear angular extent across each radio source. We identify 82 radio sources this way. For radio sources with $\theta_{\mathrm{GLASS~5.5}} < 25$'', we use the aforementioned image overlays to identify any radio sources for which the low-surface-brightness lobes escape detection in GLASS~5.5. For such cases, the angular size is measured using EMU-ES, and are accepted if $\theta_{\mathrm{EMU-ES}} \geq 25''$ (e.g. see Figure~\ref{fig:MIDAS_rest_example}). An additional 27 radio sources are identified this way, giving a total of 109 radio sources greater than 25$''$. For consistency, the angular size of each radio source is re-measured using EMU-ES, by considering the largest angular size subtended within the footprint of radio emission above 5$\sigma$.

\item \textit{AGN dominated}\\
As a result of their sensitivity to low-surface-brightness emission, both MIDAS~ES and EMU-ES are able to detect radio emission from a typical face-on spiral galaxy. Radio emission from these objects is not driven by a radio-loud AGN, and therefore these sources need to be removed from the sample. While the radio emission of virtually all radio galaxies extends well beyond the host galaxy, radio emission from spiral galaxies is associated only with the optical component of the galaxy. Thus using the aforementioned image overlays, we remove three radio sources that trace the optical/near-infrared component of the host galaxy, as revealed with VIKING $K_s-$band imaging.We provide an example in Figure~\ref{fig:MIDAS_SFG}. The remaining sample contains 106 extended radio galaxies, forming the parent sample for this analysis.

\item \textit{Activity status}\\
To constrain the nuclear activity associated with an AGN, we use GLASS~5.5 imaging to search for evidence of a radio core. For a successful radio core detection, we require a compact object with a peak flux density greater than 3$\sigma$. We use \textsc{bane} to produce an RMS image associated with each GLASS~5.5 image cutout. Only pixel values above 3$\sigma$ are considered. We use the orientation and morphology of the radio lobes as a rough constraint on the potential position of the radio core. Following this method, we classify 94 radio galaxies as active, and a further 11 as candidate remnant radio galaxies. We emphasise that this method only selects \textit{candidate} remnant radio galaxies, since the existence of a faint, undetected radio core is still possible. For each remnant candidate we place a 3$\sigma$ upper limit on the peak flux density of the core. Here, $\sigma$ is measured by drawing a circle equivalent to four GLASS synthesized beams at the position of the presumed host galaxy and measuring the RMS within this region.

\item \textit{Host identification}\\
For radio sources with a core, we use a 1$''$ search radius to cross match the position of the radio core with the GAMA~23 photometry catalog. Out of 94 such sources, the hosts of 80 radio galaxies are identified this way, of which 26 and 54 have spectroscopic and photometric redshifts, respectively. The hosts of the remaining 14 sources are either extremely faint in $K_s$-band, or remain completely undetected in VIKING, potentially due to lying at higher redshift. Host identification for remnant candidates is discussed on a per-source basis in Sect.~\ref{sec:results}, as this is a complicated and often ambiguous procedure. For each radio source we also use WISE \citep{2010AJ....140.1868W} 3.4$\mu$\,m and 4.6$\mu$\,m images to determine if any potential hosts were not present in the VIKING imaging, however this did not reveal any new candidates. 
\end{enumerate}

\subsection{Collating flux densities}
For each of the 104 radio galaxies, integrated flux densities are compiled from the data described in Sect.~\ref{sec:radiodata}. To compile the integrated flux densities at 119, 154 and 186\,MHz  (GLEAM~SGP), 216\,MHz (MIDAS~ES), and 1400\,MHz (NVSS), we use their appropriate source catalogues described in their relevant data sections. As a result of the high spatial resolution at 399\,MHz (uGMRT observations), 887\,MHz (EMU-ES) and 5.5\,GHz (GLASS~5.5), sources are often decomposed into multiple components. To ensure the integrated flux densities are measured consistently across these surveys, we convolve their image cutouts of each source with a $54''$ circular resolution (e.g. the major axis of the MIDAS synthesized beam). Integrated flux densities are then extracted using \textsc{aegean} by fitting a Gaussian to radio emission. For each remnant candidate observed with ATCA at low resolution at 2.1, 5.5 and 9\,GHz, we use \textsc{aegean} to measure their integrated flux densities. The 5.5\,GHz integrated flux density reported for each remnant candidate is exclusively taken from these low resolution ATCA observations, not GLASS. Finally, for each integrated flux density measurement, uncertainties are calculated as the quadrature sum of the measurement uncertainty and the absolute flux-scale uncertainty.

\subsection{Radio SED Fitting}
\label{sec:sed}
To better understand their energetics, we model the integrated radio spectrum of each remnant candidate. We use a standard power-law model of the radio continuum spectrum (Eqn~\ref{eqn:powlaw}), where the spectral index, $\alpha$, and the flux normalization $S_0$ are constrained by the fitting, and $\nu_0$ is the frequency at which $S_0$ is evaluated:
\begin{equation}
\label{eqn:powlaw}
    S_\nu~=~S_0~(\nu/\nu_0)^\alpha
\end{equation}
Given we can expect to see evidence of a curvature in their spectra, especially over such a large frequency range, we also fit a generic curved power-law model (Eqn~\ref{eqn:curvpowlaw}). Here, $q$ offers a parameterization of the curvature in the spectrum, where $q<0$ describes a convex spectrum. For optically-thin synchrotron emitted from radio lobes, $q$ typically ranges within $-0.2\leq q \leq 0$. Although $q$ is not physically motivated, \citet{2012MNRAS.421..108D} show that it can be related to physical quantities of the plasma lobes such as the energy and magnetic field strength.
\begin{equation}
    \label{eqn:curvpowlaw}
    S_\nu~=~S_0~\nu^\alpha e^{q(\mathrm{ln}\nu)^2}
\end{equation}
Fitting of each model is performed in Python using the \textsc{curve\_fit} module; fitted models are presented in Figures~\ref{fig:MIDAS_J225522-341807}$-$\ref{fig:MIDAS_J230442-341344}. Additionally, we calculate a Bayesian Information Criterion (BIC) for each model. To compare each model, we calculate a $\Delta$BIC~=~BIC$_{\mathrm{1}}-$BIC$_{\mathrm{2}}$ which suggests a preference towards the second model for $\Delta$BIC$>0$ (and similarly, a preference towards the first model if $\Delta$BIC$<0$). Weak model preference is implied by the following range $0<|\Delta$BIC|~$<2$, whereas a model is strongly preferred if $|\Delta$BIC|~$>6$. In Table~\ref{tab:modelledradio}, we calculate $\Delta$BIC~=~BIC$_{\mathrm{power-law}}-$BIC$_{\mathrm{curved-power-law}}$.

\begin{table*}
\centering
\caption{Summarized radio properties of the selected remnant candidates. S$_{216}$ gives the 216\,MHz integrated flux density. LAS gives the largest angular size measured from EMU-ES. S$_{\mathrm{core}}$ gives the 5.5\,GHz upper limit placed on the radio core peak flux density using GLASS. $\alpha_{\mathrm{fit}}$ denotes the spectral index fitted by each model. The curvature term modelled by the curved power-law model is represented by $q$. As per Sect.~\ref{sec:sed}, the $\Delta$BIC is calculated between each model and presented in the final column. A reduced chi-squared ($\chi^2_{\mathrm{red}}$) is also evaluated for each model.}
\label{tab:modelledradio}
\begin{tabular}{ccccc|cc|ccc|c}
\hline
\hline
\multirow{2}{*}{MIDAS Name} &Fig. &  S$_{216}$& LAS&S$_{\mathrm{core}}$ & \multicolumn{2}{c}{Power-law}  &\multicolumn{3}{c}{Curved power-law} & $\Delta$BIC\\ \cline{6-10}
&&\scriptsize{(mJy)}&\scriptsize{($''$)}&\scriptsize{($\mu$Jy\,beam$^{-1}$)} & $\alpha_{\mathrm{fit}}$ & $\chi^2_{\mathrm{red}}$ &  $\alpha_{\mathrm{fit}}$ & q & $\chi^2_{\mathrm{red}}$& \\
\hline
\hline
\footnotesize{J225522$-$341807} & \ref{fig:MIDAS_J225522-341807} &24.7$\pm$1.9&100&$<73$  & \footnotesize{-1.40$\pm$0.09} & \footnotesize{2.2} & \footnotesize{-1.50$\pm$0.10} & \footnotesize{-0.11$\pm$0.08}  & \footnotesize{1.5} & \footnotesize{-13.2}\\
\footnotesize{J225607$-$343212} & \ref{fig:MIDAS_J225607-343212} &18.3$\pm$2.0 &83&$<72$  & \footnotesize{-1.10$\pm$0.04} & \footnotesize{1.2} & \footnotesize{-1.20$\pm$0.01} & \footnotesize{-0.073$\pm$0.01} & \footnotesize{0.1} & \footnotesize{3.1} \\
\footnotesize{J225608$-$341858} & \ref{fig:MIDAS_J225608-341858} &$14.7\pm2.0$&60&$<72$   & \footnotesize{-0.86$\pm$0.05} & \footnotesize{0.3} & \footnotesize{-0.91$\pm$0.1}  & \footnotesize{-0.04$\pm$0.07}  & \footnotesize{0.3} & \footnotesize{-2.2}\\
\footnotesize{J225337$-$344745} & \ref{fig:MIDAS_J225337-344745} &$170.6\pm5.8$ &105&$<74$& \footnotesize{-0.92$\pm$0.05} & \footnotesize{12.1} & \footnotesize{-0.89$\pm$0.02} & \footnotesize{-0.12$\pm$0.02}  &\footnotesize{0.8} & \footnotesize{23.8}\\
\footnotesize{J225543$-$344047} & \ref{fig:MIDAS_J225543-344047} &$192.7\pm5.8$ &84&$<117$& \footnotesize{-0.87$\pm$0.01} & \footnotesize{0.4} & \footnotesize{-0.87$\pm$0.02} & \footnotesize{-0.003$\pm$0.01} & \footnotesize{0.5}& \footnotesize{-2.8}\\
\footnotesize{J225919$-$331159} & \ref{fig:MIDAS_J225919-331159} &$36.6\pm2.0$&70&$<60$   & \footnotesize{-0.86$\pm$0.02} & \footnotesize{1.0} & \footnotesize{-0.89$\pm$0.02} & \footnotesize{0.035$\pm$0.01}  &\footnotesize{0.5} & \footnotesize{-6.9}\\
\footnotesize{J230054$-$340118} & \ref{fig:MIDAS_J230054-340118} &$113.4\pm6.7$&106&$<86$ & \footnotesize{-0.73$\pm$0.03} & \footnotesize{2.8} & \footnotesize{-0.72$\pm$0.03} & \footnotesize{-0.027$\pm$0.02} & \footnotesize{2.5}& \footnotesize{-4.1}\\
\footnotesize{J230104$-$334939} & \ref{fig:MIDAS_J230104-334939} &$55.4\pm2.2$&37&$<57$   & \footnotesize{-0.72$\pm$0.03} & \footnotesize{2.0} & \footnotesize{-0.69$\pm$0.04} & \footnotesize{-0.023$\pm$0.01} & \footnotesize{2.0}& \footnotesize{2}\\
\footnotesize{J230321$-$325356} & \ref{fig:MIDAS_J230321-325356} & $153.6\pm6.0$&93&$<74$ & \footnotesize{-0.86$\pm$0.01} & \footnotesize{0.9} & \footnotesize{-0.85$\pm$0.01} & \footnotesize{-0.019$\pm$0.01} & \footnotesize{0.5}& \footnotesize{6.4}\\
\footnotesize{J230442$-$341344} & \ref{fig:MIDAS_J230442-341344} &$198.1\pm6.1$&50&$<84$  & \footnotesize{-1.00$\pm$0.02} & \footnotesize{1.2} & \footnotesize{-1$\pm$0.02}    & \footnotesize{-0.008$\pm$0.02} &\footnotesize{1.5} & \footnotesize{-1.3}\\
\hline
\end{tabular}
\end{table*}

\section{Remnant radio galaxy candidates}
\label{sec:results}
We present and discuss each of the 11 candidate remnant radio galaxies below. Seven are found to display hotspots in GLASS. Image overlays and the radio continuum spectrum are presented in Figure~\ref{fig:targets}. General radio properties are presented in Table~\ref{tab:modelledradio}.
\subsection{Remnant candidates without hotspots}
\label{sec:remnants_without_hotspots}
\subsubsection{MIDAS J225522-341807}
\label{sec:MIDAS_J225522-341807}
\textit{Radio properties.}
Figure \ref{fig:MIDAS_J225522-341807} shows extremely relaxed lobes and an amorphous radio morphology. No compact structures that would indicate hotspots are observed. The average 154\,MHz surface brightness is $\sim32\,$mJy\,arcmin$^{-2}$, satisfying the low SB criterion (SB~$< 50$\,mJy\,arcmin$^{-2}$) employed by \citet{2017A&A...606A..98B}. The diffuse radio emission is undetected by the uGMRT observations, NVSS, GLASS, as well as the 2.1\,GHz and 9\,GHz ATCA follow-up observations. Unsurprisingly, we find that the source spectrum appears ultra-steep at low frequencies, and demonstrates a curvature ($q=-0.11$) across the observed range of frequencies. The radio properties point towards an aged remnant. 

\textit{Host galaxy.} Identification of the host galaxy is rather challenging here as the amorphous radio morphology provides little constraints on the host position. No clear host galaxy is seen along the centre of the radio emission, however this can easily be explained if the radio lobes have drifted. We approximate the central position of the radio emission by taking the centre of an ellipse drawn to best describe the radio source. G1~($z_p=0.474$) is located 10.2$''$ from the radio center, corresponding to a 61\,kpc offset. G2~($z_p=0.433$) is located 14.1$''$ from the radio center, corresponding to a 80\,kpc offset. G3~($z_p=0.294$) is located 23$''$ from the radio center, corresponding to a 102\,kpc offset. Without any additional information, we take G1 as the likely host galaxy. We note that G4~($z_p=0.41$) shows compact radio emission at 887\,MHz, however it is unclear whether this is related to the extended structure. We include the radio spectrum arising from G4 in Figure~\ref{fig:MIDAS_J225522-341807}, and note that it contributes approximately 5\% to the total radio flux density at 887\,MHz. If G4 is unrelated, its radio spectrum should be subtracted from the integrated spectrum of MIDAS~J225522-341807.

\subsubsection{MIDAS J225607-343212}
\label{sec:MIDAS_J225607-343212}
\textit{Radio properties.} Figure \ref{fig:MIDAS_J225607-343212} shows a pair of relaxed radio lobes, with a diffuse bridge of emission connecting each lobe along the jet axis. The 154\,MHz average surface brightness is calculated as $\sim26$mJy\,arc-minute$^{-2}$, satisfying the low surface brightness criterion. The edge brightened regions likely represent the expanded hotspots of the previously active jet, similar to what is observed in B2~0924+30 \citep{2017A&A...600A..65S}. The source is undetected by the uGMRT observations, GLASS, as well as the ATCA follow-up at 2.1\,GHz and 9\,GHz. Curvature is evident in the spectrum, which becomes ultra-steep above 1.4\,GHz.

\textit{Host galaxy.} Along the projected centre of the jet axis, a collection of three potential host galaxies exist within a $\sim7''$ aperture. The redshift of each host, G1~($z_s=0.31307$), G2~($z_p=0.361$), G3~($z_s=0.27867$), suggests they are all at similar redshift, and thus would not result in an appreciable difference to the corresponding physical size and radio power. We take G1 as the most likely host as it lies closest to the projected centre of the radio lobes. 
\subsubsection{MIDAS J225608-341858}
\label{sec:MIDAS_J225608-341858}
\textit{Radio properties.} Figure \ref{fig:MIDAS_J225608-341858} shows two relaxed, low surface brightness lobes that are asymmetrical in shape. The flattened `pancake'-like morphology of the Northern lobe can be explained by the buoyant rising of the lobes \citep{2001ApJ...554..261C}. The surface brightness of each lobes is approximately $43$\,mJy\,arcmin$^{-2}$, satisfying the low SB criterion employed by \citet{2017A&A...606A..98B}. The source is undetected by the follow-up 2.1\,GHz and 9\,GHz ATCA observations. The spectrum seems consistent with a single power law ($\alpha=-0.86$), and the detection at 5\,GHz is too weak to determine whether spectral curvature is evident at higher frequencies. 

\textit{Host galaxy.} G1~($z_p=0.57$) lies $4.8''$ away from the radio center, corresponding to a 32\,kpc offset. G2~($z_p=0.321$) lies $13''$ away from the radio center, corresponding to a 61\,kpc offset. This assumes that the lobes are equidistant from the host, which is not always the case. However, we retain G1 as the likely host galaxy. 
\subsection{Candidates with hotspots}
\subsubsection{MIDAS J225337-344745}
\label{sec:MIDAS_J225337-344745}
\textit{Radio properties.} Figure \ref{fig:MIDAS_J225337-344745} shows a typical low-resolution FR-II radio galaxy as evidenced by the edge-brightened morphology. The average 154\,MHz surface brightness is $160$\,mJy~arcmin$^{-2}$. The source is firmly detected by the ATCA follow-up at all frequencies, revealing that the spectrum is highly curved ($q=-0.12$) over the observed frequency range and only becomes ultra-steep at $\nu \gtrsim 2$\,GHz. An evaluation of its spectral curvature reveals SPC~$= 0.78 \pm 0.17$, suggesting the lobes are remnant. The properties of the spectrum strongly suggest a lack of energy supply to the lobes, however, GLASS~5.5 reveals compact emitting regions at the edges of each lobe that may suggest recent energy injection. We divert a detailed analysis of this source to Sect.~\ref{sec:evolutionary}.

\textit{Host galaxy.} The radio lobes are unambiguously associated with galaxy G1 ($z_s=0.2133$) based on the close proximity to the geometric centre of the lobes.
\subsubsection{MIDAS J225543-344047}
\label{sec:MIDAS_J225543-344047}
\textit{Radio properties.} Figure \ref{fig:MIDAS_J225543-344047} demonstrates an elongated, `pencil-thin', radio galaxy with an edge-brightened FR-II morphology. GLASS detects only the brightest and most compact emitting regions, and misses the lower surface brightness emission seen at 887\,MHz. The radio source is detected in all but the 2.1\,GHz ATCA follow-up observations. The radio spectrum is well modelled by a power-law ($\alpha=-0.87$), and shows no evidence of a curvature up to 9\,GHz. 

\textit{Host galaxy.} G1~($z_p=1.054$) lies closest to the projected radio center. G2~($z_p=1.019$) and G3~($z_p=1.342$) are also likely, however the difference in implied redshift would not result in an appreciable change to the derived physical size and radio power.
\subsubsection{MIDAS J225919-331159}
\label{sec:MIDAS_J225919-331159}
\textit{Radio properties.} Figure \ref{fig:MIDAS_J225919-331159} demonstrates a pair of lobes with compact emitting regions seen by GLASS. The radio spectrum is well approximated as a power-law ($\alpha=-0.86$) with no evidence of a spectral curvature over the observed range of frequencies. 

\textit{Host galaxy.} G1~($z_p=0.504$) is taken as the likely host galaxy due to its central position between each lobe.
\subsubsection{MIDAS J230054-340118}
\label{sec:MIDAS_J230054-340118}
\textit{Radio properties.} Figure \ref{fig:MIDAS_J230054-340118} shows a peculiar radio morphology; while the western lobe shows bright emitting regions in GLASS, the counter lobe is completely diffuse and does not show a hotspot. It is unclear what is causing this. 

\textit{Host galaxy.} The galaxies G1~($z_p=0.32$) and G2~($z_s=0.306$) are considered as host candidates, given their position between each lobe. We find that G2 is not associated with any galaxy group; if the lobe asymmetry is due to an environmental effect, we would not expect a field galaxy host. We can not comment on whether G1 is associated with a group, given the lack of a spectroscopic confirmation.

\subsubsection{MIDAS J230104-334939}
\label{sec:MIDAS_J230104-334939}
\textit{Radio properties.} Figure \ref{fig:MIDAS_J230104-334939} shows a typical FR-II radio galaxy, as implied by the edge-brightened morphology. The source exhibits clear hotspots in each lobe, as seen by GLASS~5.5 and GLASS~9.5. The source is detected in all but the 2.1\,GHz ATCA follow-up. Modelling the radio continuum spectrum gives a spectral index of approximately $\alpha=-0.7$, revealing no significant energy losses. The $\Delta$BIC offers tentative evidence for some spectral curvature, however this may just be a result of a poorly constrained spectrum at low ($\leq215$\,MHz) frequencies.

\textit{Host galaxy.} The radio source is unambiguously associated with G1 $(z_s=0.312)$, which almost perfectly aligns with the projected centre of the source.

\begin{figure*}
    \begin{subfigure}{1\linewidth}
     \caption{\textbf{MIDAS J225522$-$341807}. EMU-ES contour levels:~[3,4,5,7,10]$\times\sigma$. GLASS~5.5 contour levels:~[3,4,5]$\times\sigma$. GLASS~9.5 contours are not presented due to an absence of radio emission above $3\sigma$. Compact component at RA=22$^\mathrm{h}$55$^\mathrm{m}$25.5$^\mathrm{s}$, Dec= -34$^\circ$18$'$40$''$ is unrelated. The radio spectrum of the compact radio component, G4, is demonstrated by the blue markers. Radio emission from G4 is undetected by GLASS~5.5, we thus present a 3$\sigma$ upper limit.}
    \centering\includegraphics[width=\linewidth]{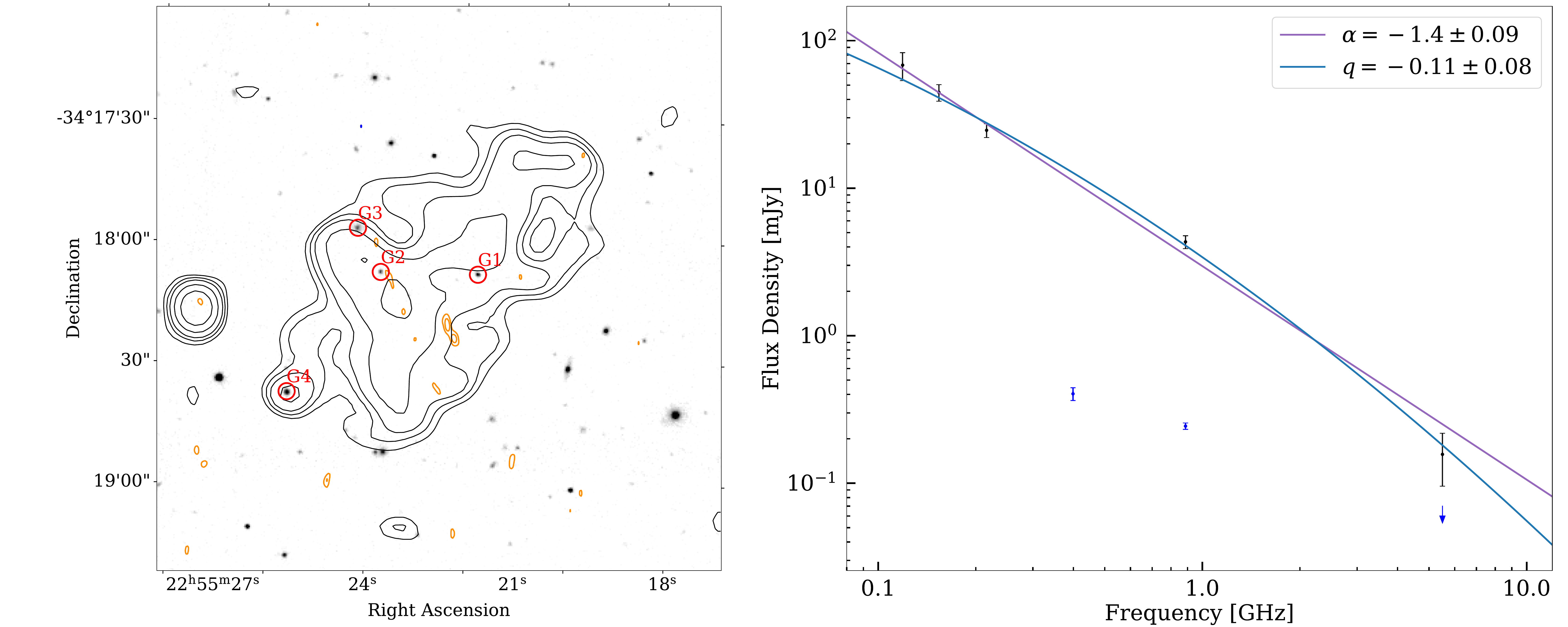}
    \label{fig:MIDAS_J225522-341807}
    \end{subfigure}\hfill
    \centering
    \begin{subfigure}{1\linewidth}
    \caption{\textbf{MIDAS J225607$-$343212}. EMU-ES contour levels:~[3,4,5,7,10,12,15,20]$\times\sigma$. GLASS~5.5 contour levels:~[3,4,5]$\times\sigma$. GLASS~9.5 contours are not presented due to an absence of radio emission above $3\sigma$. Compact component at RA=22$^\mathrm{h}$56$^\mathrm{m}$03$^\mathrm{s}$, Dec= -34$^\circ$32$'$55$''$ is unrelated. }
    \centering\includegraphics[width=\linewidth]{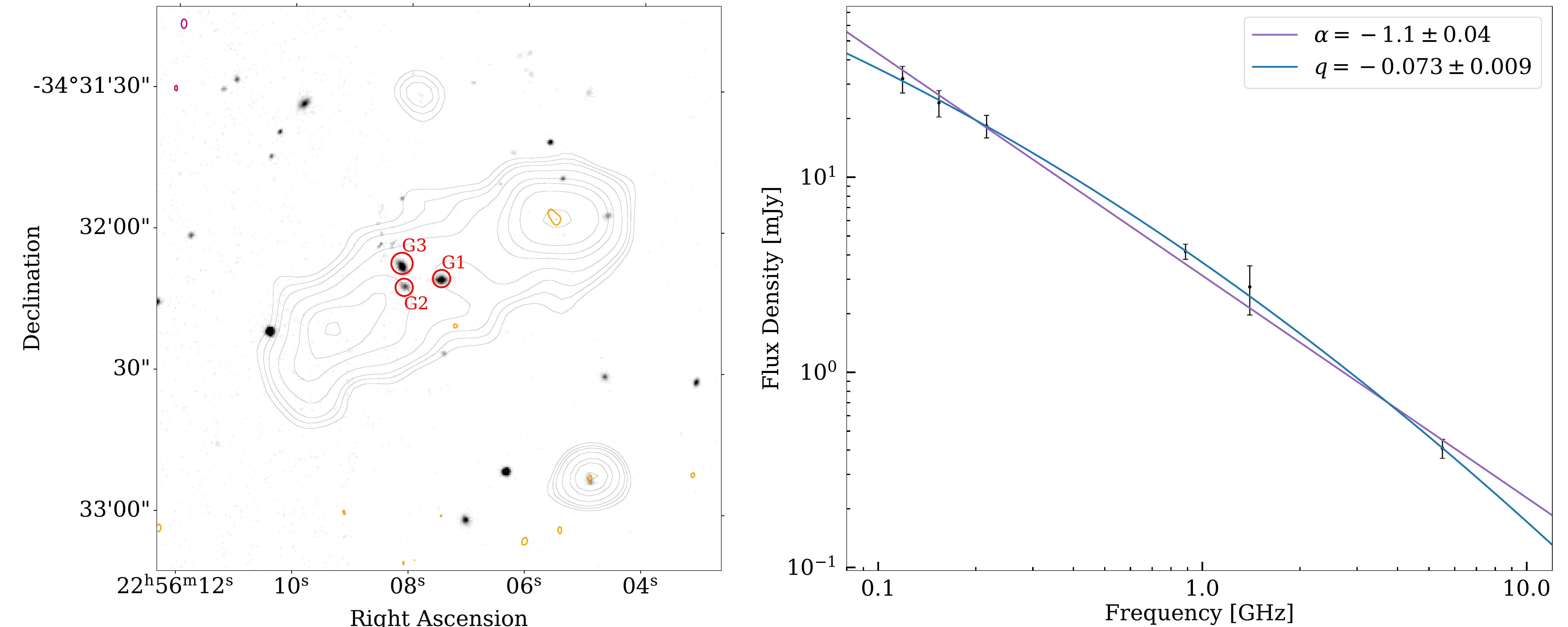}
    \label{fig:MIDAS_J225607-343212}
    \end{subfigure}
    \caption{\textbf{Left:} Plotted are the remnant candidates presented in Sect.~\ref{sec:results}. Background image is a VIKING K$_s$ band cutout set on a linear stretch. Three sets of contours are overlaid, representing the radio emission as seen by EMU-ES \textit{(black)}, GLASS~5.5 \textit{(orange)} and GLASS~9.5 \textit{(blue)}. Red markers are overlaid on the positions of potential host galaxies. \textbf{Right:} The radio continuum spectrum between 119\,MHz and 9\,GHz. 
    The integrated flux densities at 5.5\,GHz come from the low-resolution ATCA observations (Sect.~\ref{sec:LSCX}) not the lower resolution GLASS images. A simple power-law (Eqn~\ref{eqn:powlaw}) and curved power-law (Eqn~\ref{eqn:curvpowlaw}) model are fit to the spectrum, indicated by the \textit{purple} and \textit{blue} models, respectively.}
    \label{fig:targets}
    \end{figure*}
    \begin{figure*}\ContinuedFloat
    \begin{subfigure}{0.9\linewidth}
     \caption{\textbf{MIDAS J225608$-$341858}. EMU-ES contour levels:~[3,4,5,7,10]$\times\sigma$. GLASS~5.5 contour levels:~[3,4,5]$\times\sigma$. GLASS~9.5 contours are not presented due to an absence of radio emission above $3\sigma$.}
    \centering\includegraphics[width=\linewidth]{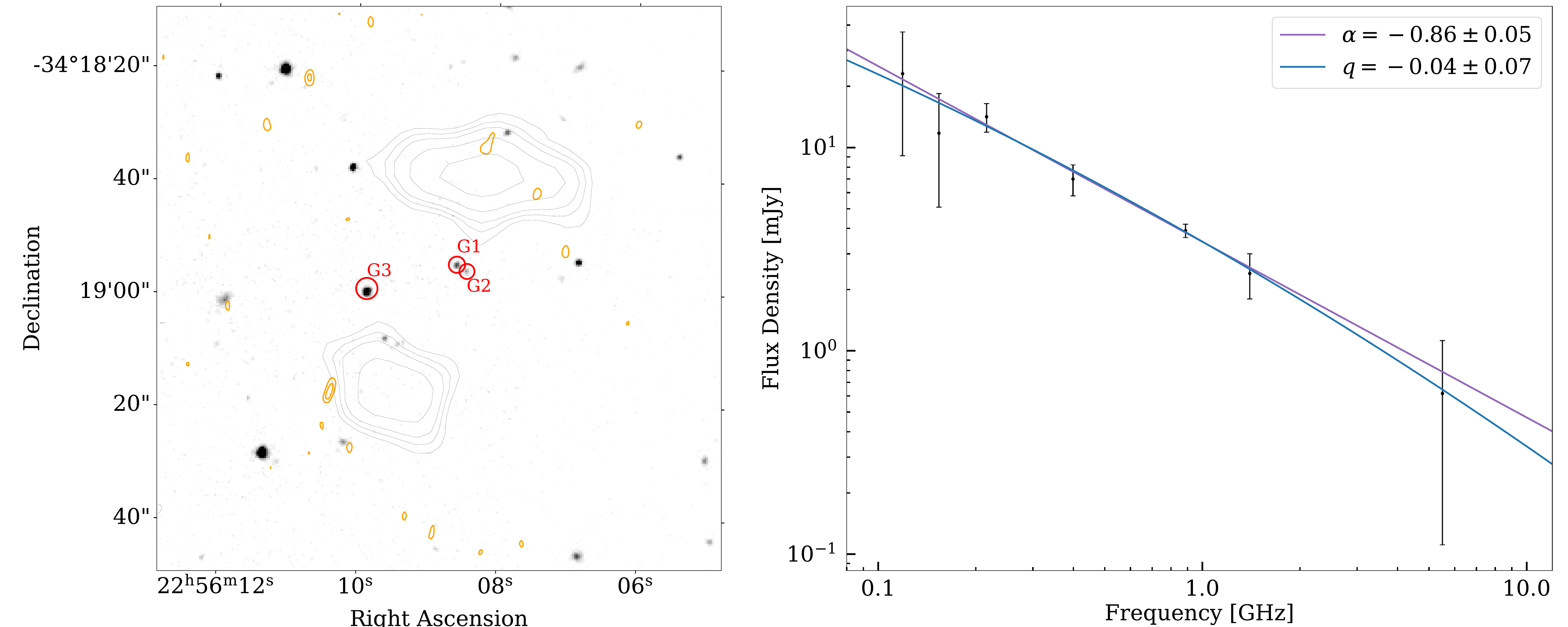}
    \label{fig:MIDAS_J225608-341858}
    \end{subfigure}
    \begin{subfigure}{0.9\linewidth}
    \caption{\textbf{MIDAS J225337$-$344745}. EMU-ES contour levels:~[4,5,10,30,50,70]$\times \sigma$, GLASS~5.5 contour levels:~[3,4,5,6]$\times \sigma$. GLASS~9.5 contours are not presented due to an absence of radio emission above $3\sigma$.}
    \centering\includegraphics[width=\linewidth]{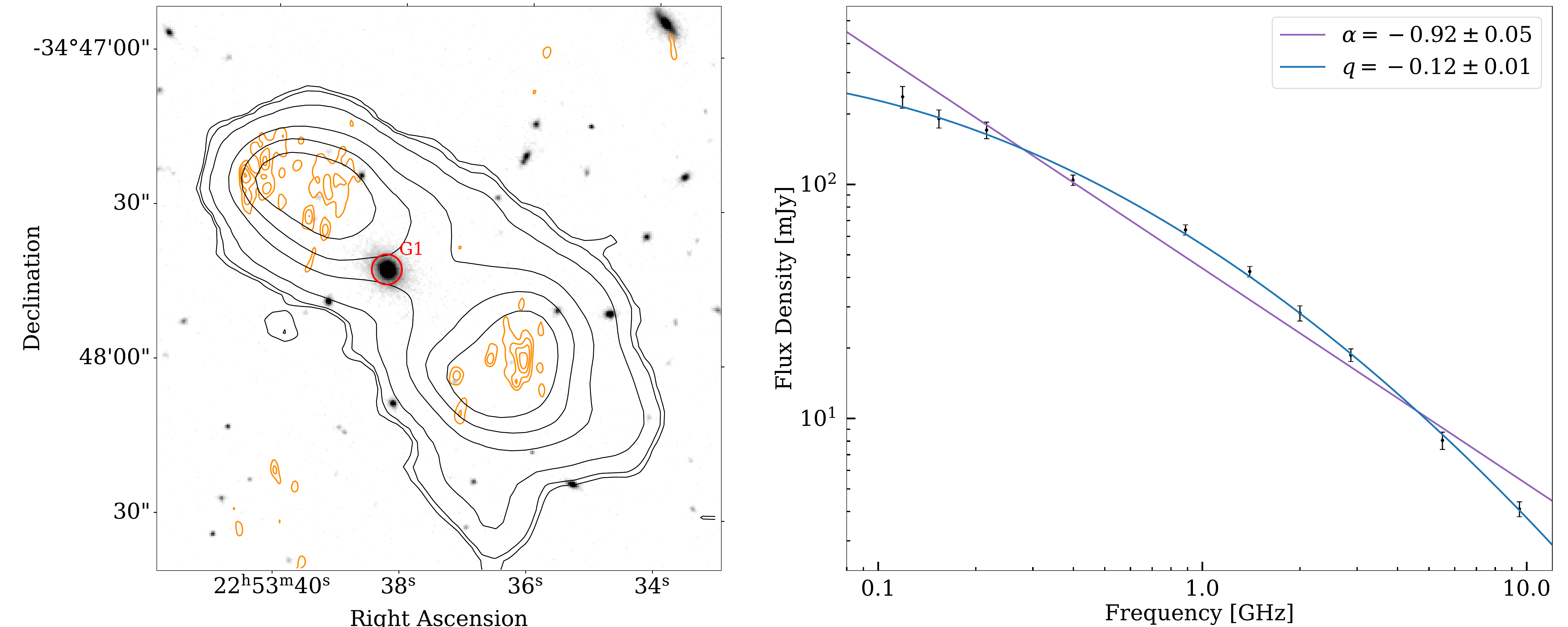}
    \label{fig:MIDAS_J225337-344745}
    \end{subfigure}
    \begin{subfigure}{0.9\linewidth}
    \caption{\textbf{MIDAS J225543$-$344047}. EMU-ES contour levels:~[3,4,5,7,15,30,100]$\times \sigma$, GLASS~5.5 contour levels:~[3,5,10,20]$\times \sigma$. GLASS~9.5 contour levels:~[3,5,10,20]$\times \sigma$}
    \centering\includegraphics[width=\linewidth]{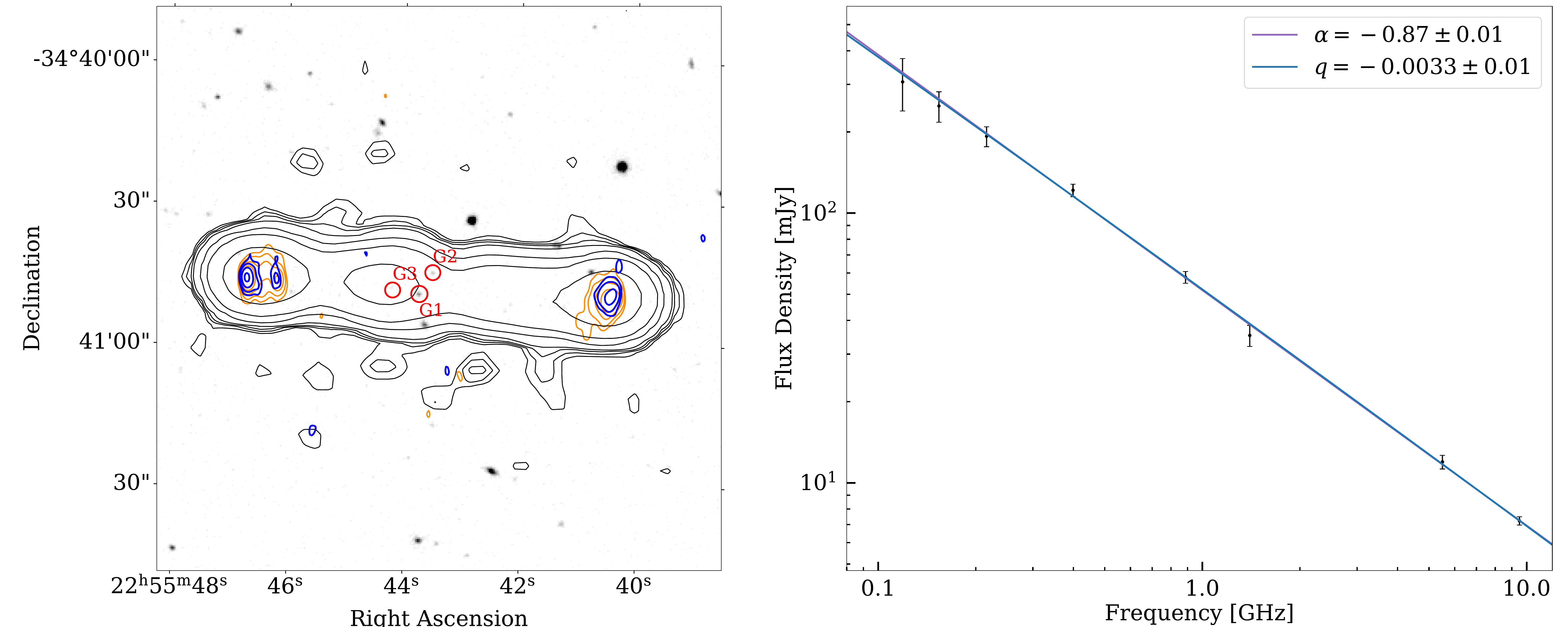}
    \label{fig:MIDAS_J225543-344047}
    \end{subfigure}
    \caption{ -- continued.}
    \end{figure*}
    \begin{figure*}\ContinuedFloat
    \begin{subfigure}{0.9\linewidth}
     \caption{\textbf{MIDAS J225919-331159}. EMU-ES contour levels:~[5,10,20,40,60]$\times \sigma$, GLASS~5.5 contour levels:~[3,4,5,6,10,20]$\times \sigma$. GLASS~9.5 contour levels:~[3,4,5,6]$\times \sigma$}
    \centering\includegraphics[width=\linewidth]{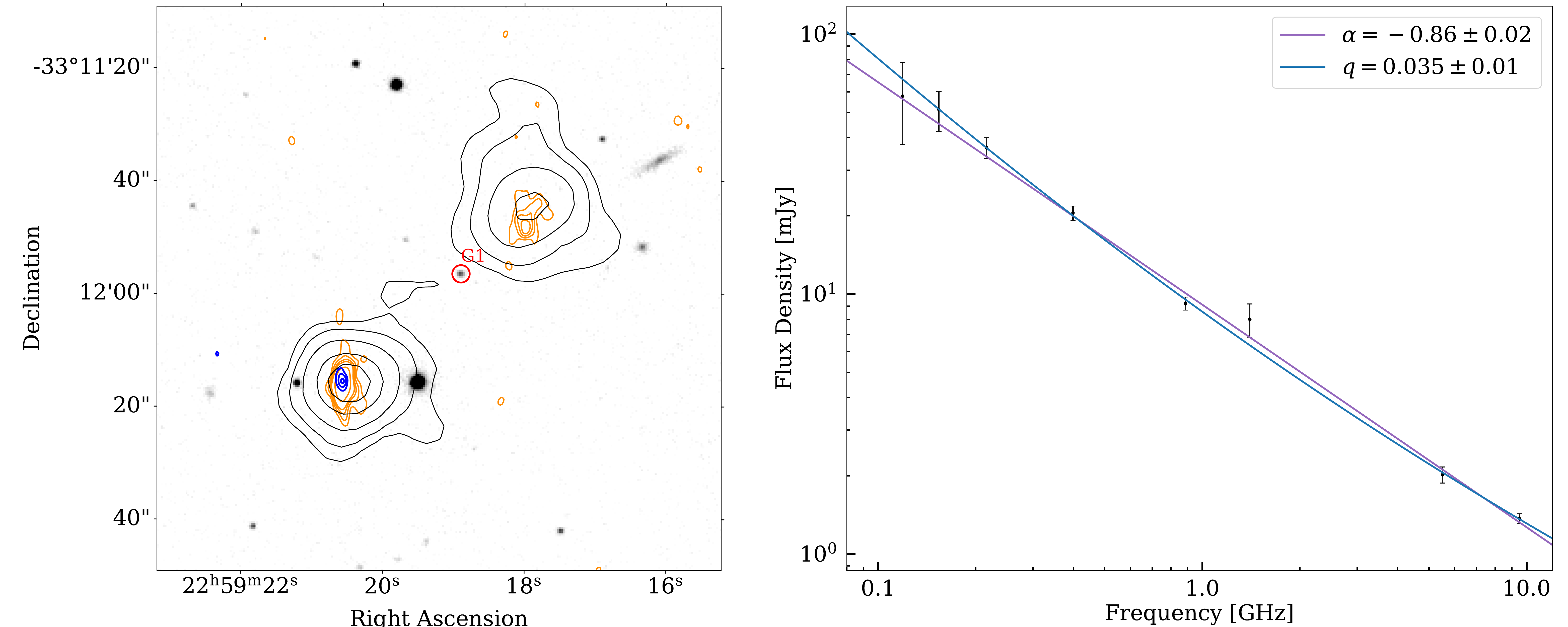}
    \label{fig:MIDAS_J225919-331159}
    \end{subfigure}\hfill
    \begin{subfigure}{0.9\linewidth}
    \caption{\textbf{MIDAS J230054$-$340118}. EMU-ES contour levels:~[3,4,5,7,15,30,100,300]$\times \sigma$, GLASS~5.5 contour levels:~[3,5,10,20,30]$\times \sigma$. GLASS~9.5 contour levels:~[3,5,10,20]$\times \sigma$}
    \centering\includegraphics[width=\linewidth]{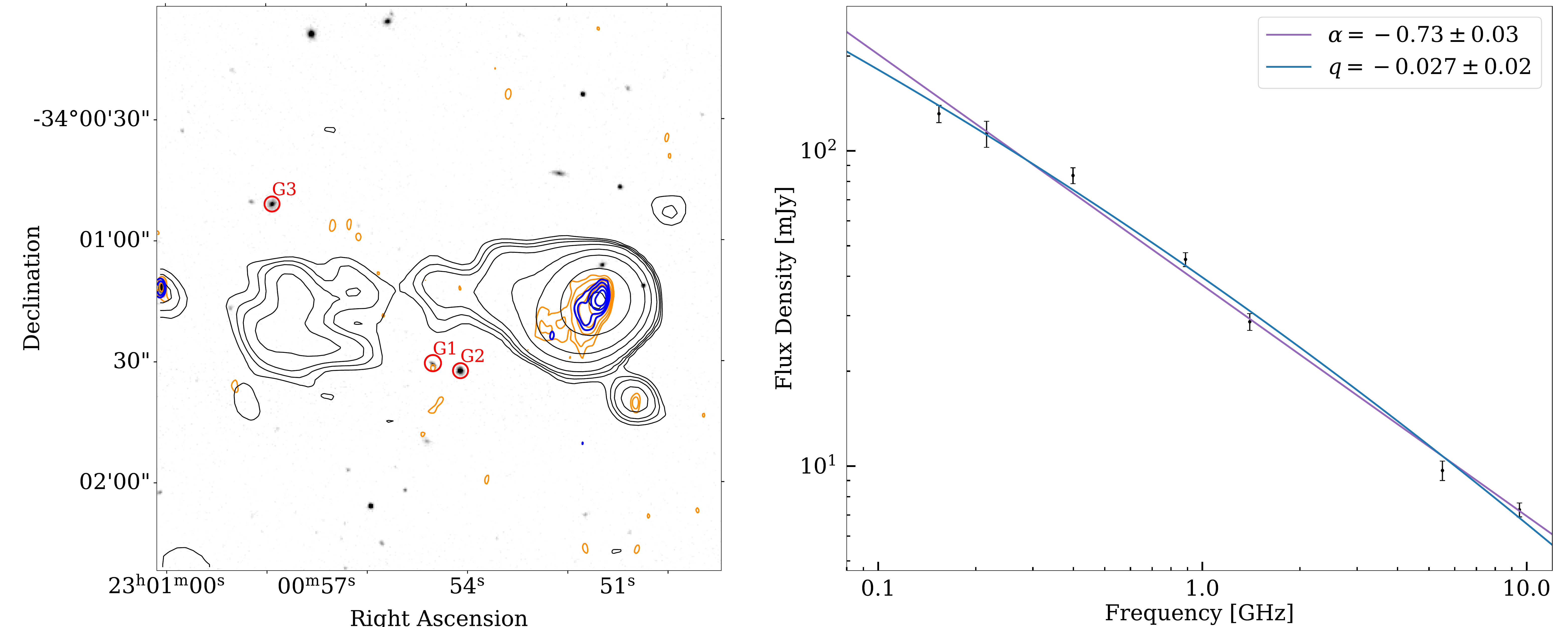}
    \label{fig:MIDAS_J230054-340118}
    \end{subfigure}
    \begin{subfigure}{0.9\linewidth}
    \caption{\textbf{MIDAS J230104-334939}. EMU-ES contour levels:~[5,8,15,35,50]$\times \sigma$, GLASS~5.5 contour levels:~[3,5,7,9,11]$\times \sigma$. GLASS~9.5 contour levels:~[3,4,5,6]$\times \sigma$}
    \centering\includegraphics[width=\linewidth]{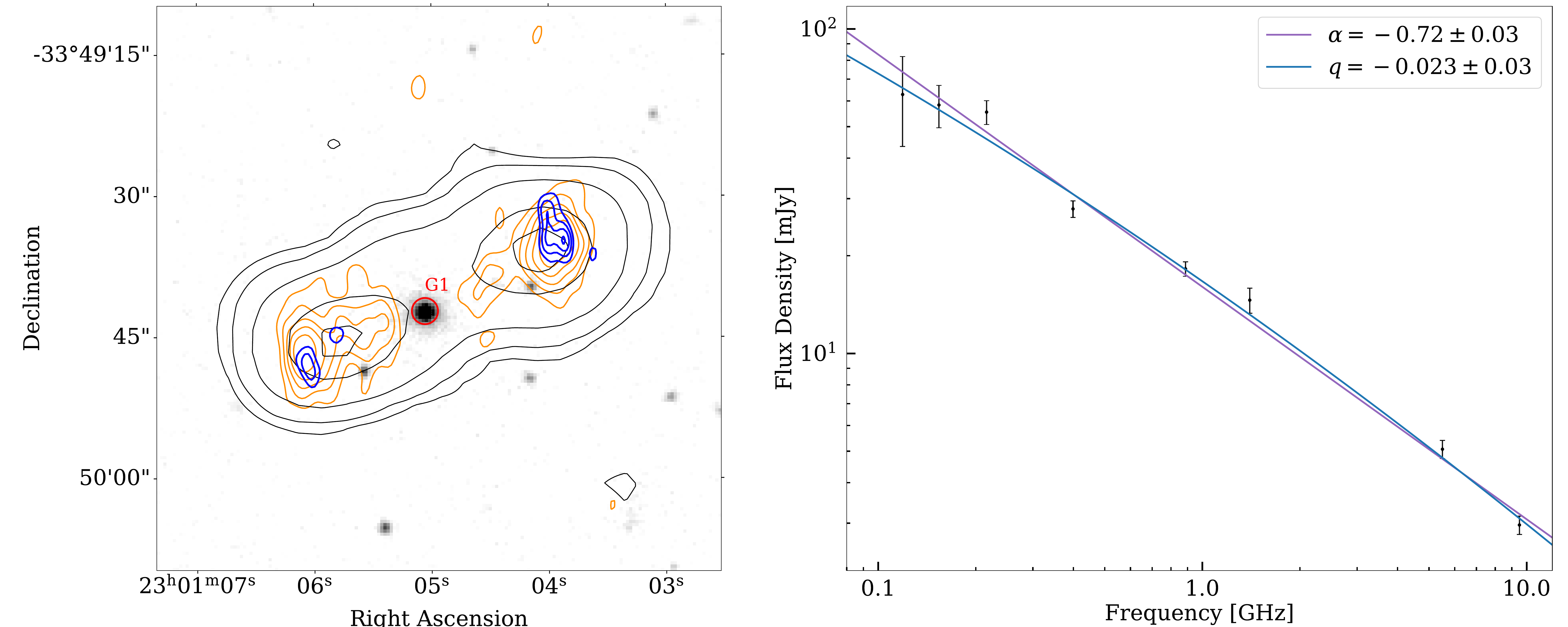}
    \label{fig:MIDAS_J230104-334939}
    \end{subfigure}\hfill
        \caption{-- continued.}
    \end{figure*}
    \begin{figure*}\ContinuedFloat
    \begin{subfigure}{\linewidth}
    \caption{\textbf{MIDAS J230321$-$325356}. EMU-ES contour levels:~[5,10,30,100,300]$\times \sigma$, GLASS~5.5 contour levels:~[3,5,10,20,30,40,50]$\times \sigma$. GLASS~9.5 contour levels:~[3,5,10,20]$\times \sigma$}
    \centering\includegraphics[width=\linewidth]{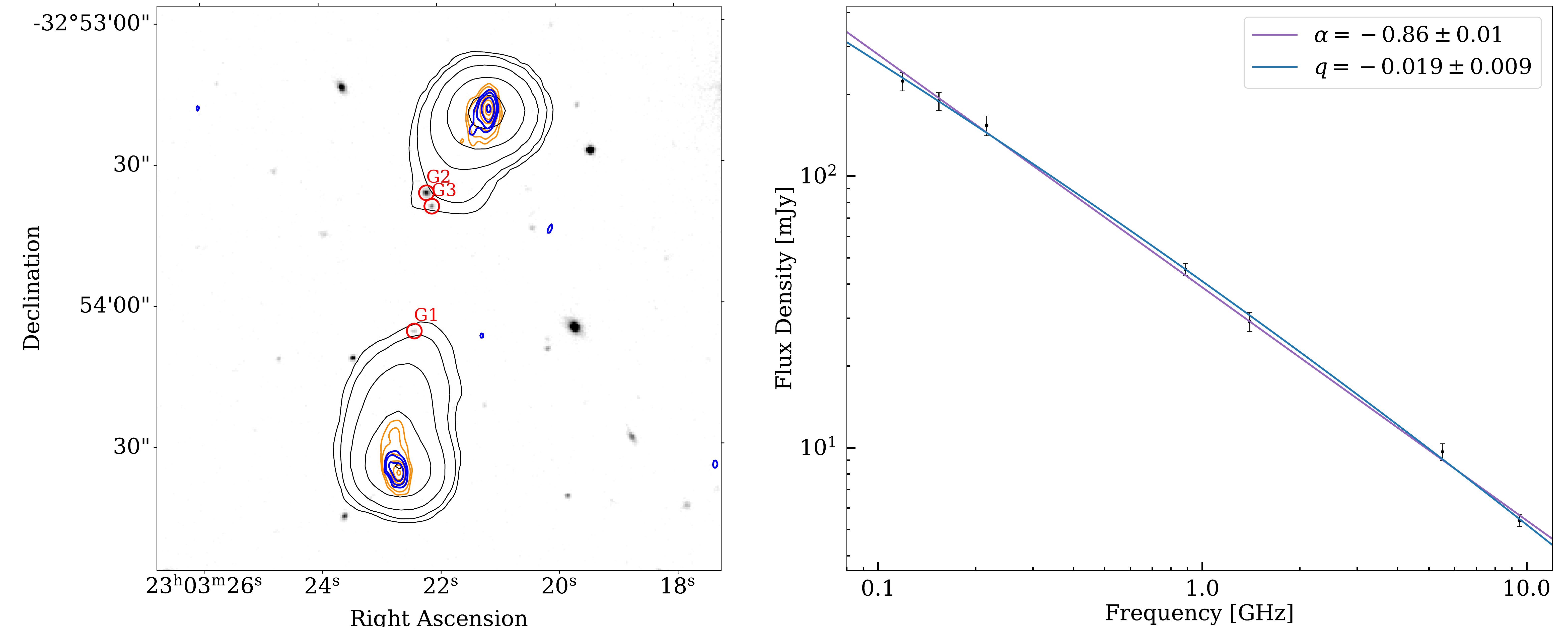}
    \label{fig:MIDAS_J230321-325356}
    \end{subfigure}\hfill
    \begin{subfigure}{\linewidth}
     \caption{\textbf{MIDAS J230442$-$341344}. EMU-ES contour levels:~[5,10,30,100,300]$\times \sigma$, GLASS~5.5 contour levels:~[3,5,10,20,30,40,50]$\times \sigma$. GLASS~9.5 contour levels:~[3,5,10,20]$\times \sigma$}
    \centering\includegraphics[width=\linewidth]{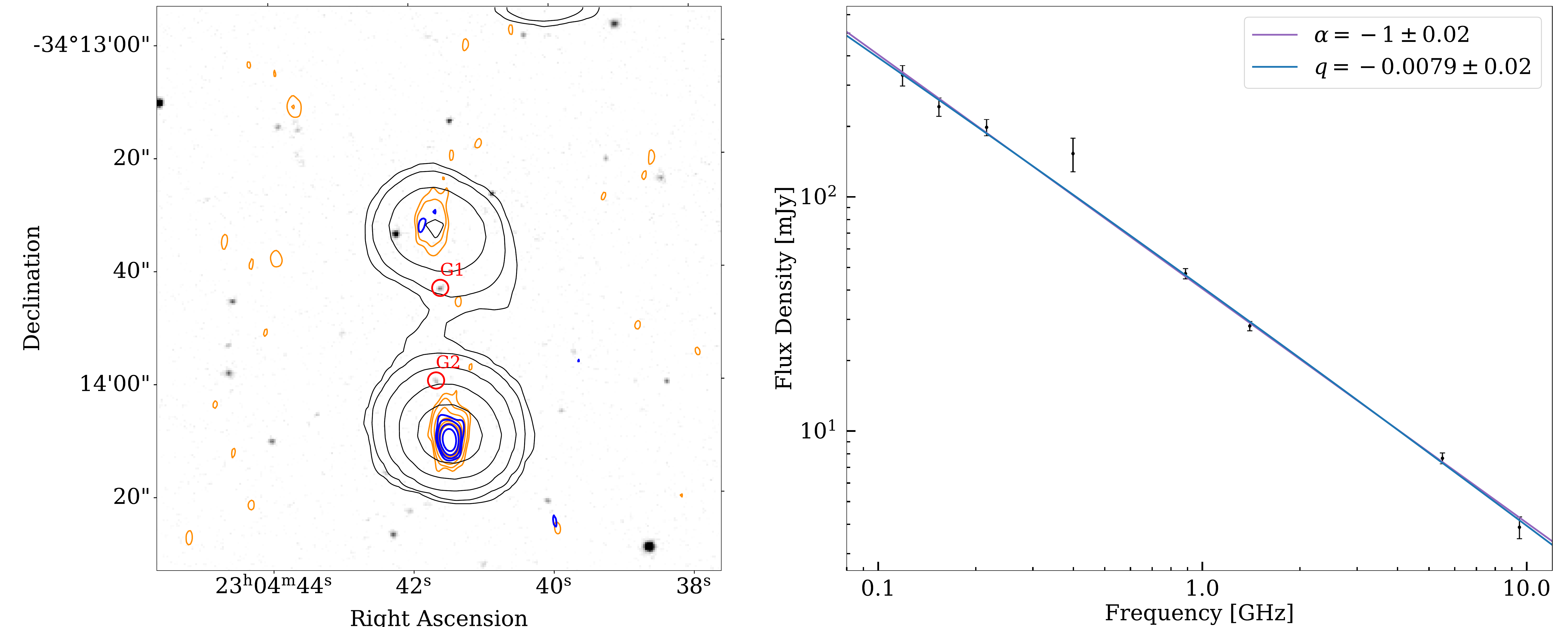}
    \label{fig:MIDAS_J230442-341344}
    \end{subfigure}
    \caption{ -- continued.}
\end{figure*}
\subsubsection{MIDAS J230321-325356}
\label{sec:MIDAS_J230321-325356}
\textit{Radio properties.} Figure \ref{fig:MIDAS_J230321-325356} shows two distinct radio lobes with compact emitting regions observed at 5.5\,GHz and 9.5\,GHz. The spectral index is approximately $\alpha=-0.86$ showing no significant evidence of spectral ageing. The $\Delta$BIC does indicate a preference towards the curved power law model, however any real curvature in the spectrum is marginal ($q\sim-0.02$) and does not necessarily require an absence of energy injection.
\textit{Host galaxy.} Three likely host galaxies are identified by their alignment along the projected jet axis; G1~($z_p=1.327$), G2~($z_p=0.622$) and G3~($z_p=0.775$). We assume G1 is the true host, due to its smaller separation from the projected radio center. 
\subsubsection{MIDAS J230442-341344}
\label{sec:MIDAS_J230442-341344}
\textit{Radio properties.} Figure \ref{fig:MIDAS_J230442-341344} . Both lobes are detected in GLASS~5.5, however, only the southern lobe shows emission in GLASS~9.5. It is unclear what is causing this; both components are resolved by GLASS and show similar morphologies, so it is unlikely they are unrelated. The radio spectrum is well approximated by a power-law ($\alpha=-1$), with no evidence of a spectral curvature.
\textit{Host galaxy.} G1~($z_p=0.9$) is a favourable host galaxy candidate, due to its small angular separation from the projected centre between the lobes. G2~($z_p=0.8075$) is located further off centre towards the South, at a similar implied redshift. 
\section{Discussion}
\label{sec:discussion}
\subsection{Sample properties}
\subsubsection{Core prominence distribution}
\label{sec:cp}
\begin{figure}
    \centering
    \includegraphics[width=0.99\linewidth]{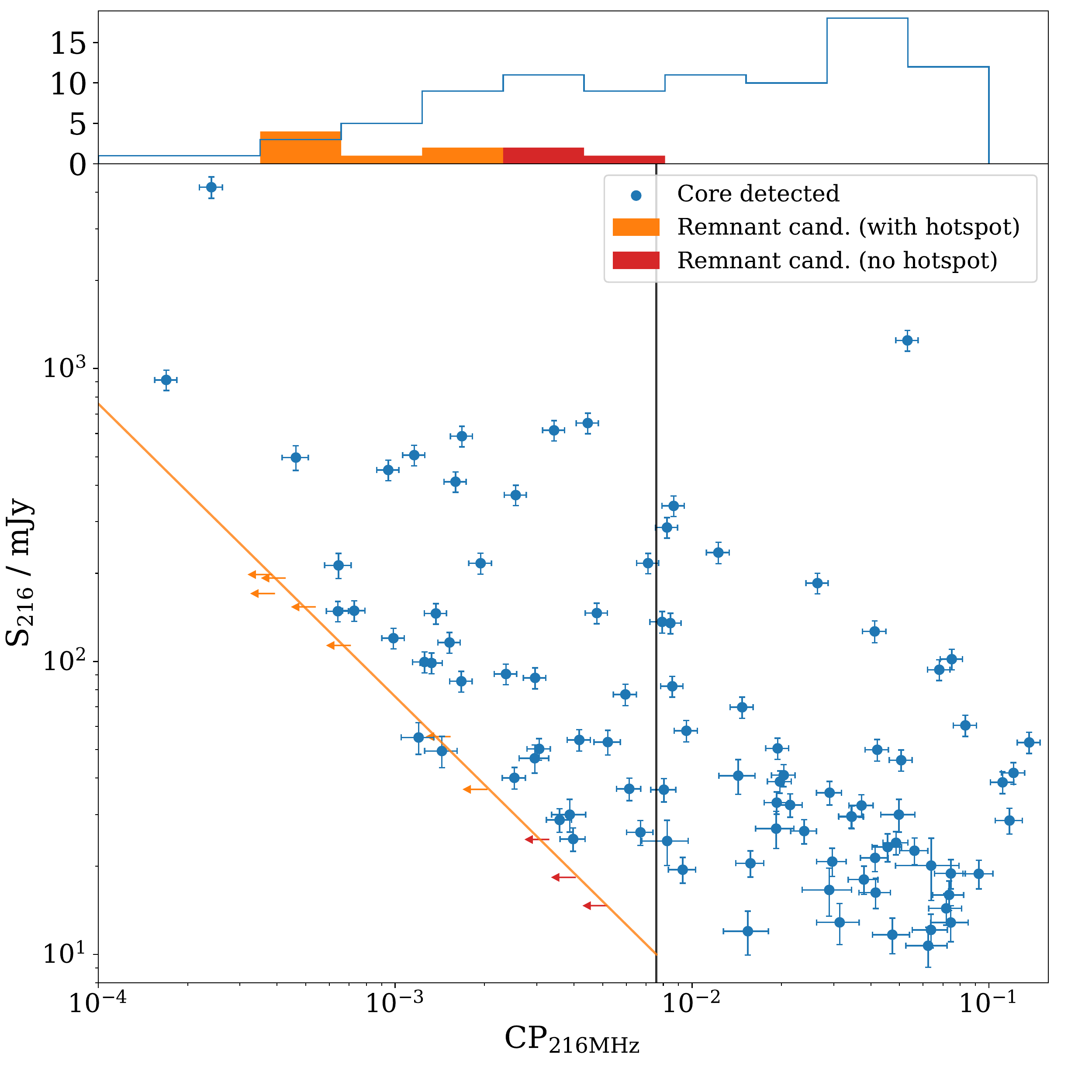}
    \caption{216\,MHz CP distribution of radio sources (see Sect.~\ref{sec:cp}). Core-detected radio galaxies are represented by the blue markers. 3$\sigma$ upper limits are placed on the remnant CP, denoted by the left-pointing arrows. Orange and red colored arrows are used to indicate remnant candidates with and without hotspots, respectively. The solid black line gives the value of the CP above which we are complete, given the 10\,mJy integrated flux density threshold and the 75$\mu$Jy beam$^{-1}$ average GLASS~5.5 detection limit. The orange line traces the lowest CP that can be recovered at the corresponding total flux density. Uncertainties on the CP are propagated from the uncertainties on the total and core flux density.} A histogram of CP is presented in the top panel.
    \label{fig:cp}
\end{figure}

To understand the limitations imposed by our selection criteria, we investigate the core prominence distribution across our sample. We define the core prominence (CP) as the ratio of core to total flux density, i.e. $\mathrm{CP} = S_{\mathrm{core}}/S_{\mathrm{total}}$. The total integrated flux density is measured at 216\,MHz. We take the GLASS~5.5 measurement of the radio core flux density, and re-scale to 216\,MHz assuming the spectrum of the radio core is a flat spectrum ($\alpha = 0$) \citep[e.g;][]{2008MNRAS.388..176H}. For the selected remnant candidates, we present upper limits on their CP by using the 3$\sigma$ upper limits on their core peak flux density. Our results are presented in Figure~\ref{fig:cp}.\\

Sources with radio cores show a wide distribution in their CP, varying within the range 10$^{-1}$~--~10$^{-4}$. The median CP of the sample is $\sim1\times10^{-2}$, almost two orders of magnitude larger than the median CP reported by \citet{2008MNRAS.390..595M} for the 3CRR sample (e.g. $\sim3\times10^{-4}$). This can be expected given the most powerful radio galaxies are preferentially selected by 3CRR. Instead, comparing our CP range to the LOFAR-selected sample compiled by \citet{2018MNRAS.475.4557M} -- e.g. see their Figure~4 -- we find the ranges are consistent. Although the reader should note \citet{2018MNRAS.475.4557M} compute their CP at 150\,MHz, meaning a $\Delta \nu=66$\,MHz frequency shift should be accounted for if a direct comparison is made. \\

As discussed in Sect.~\ref{sec:remevolution}, the `absent radio core' criterion only selects remnant candidates. Genuine remnant radio galaxies will not display a radio core, meaning their CP should approach null. In fact, the CP in such sources should be lower than any active radio source in which a radio core is present. This implies that a clean separation should exist between active and remnant radio galaxies, however, this is not what we see in Figure~\ref{fig:cp}. Instead, we find that the CP upper limits imposed on the remnant candidates overlap with core-detected radio galaxies. This comes as a result of our sample criteria. Given the GLASS detection limit ($\sim75\mu$Jy\,beam$^{-1}$), only remnant candidates brighter than $\sim500$\,mJy will show CP upper limits below what is observed for core-detected radio galaxies (e.g. log(CP)~$\lesssim -3.7$). This result indicates that it is still possible for some of the selected remnant candidates to display a faint radio core that is missed by GLASS. For remnant candidates without hotspots this is less of a concern, as we have additional information that would suggest the jets have switched off. However, for those with hotspots, decreasing the upper limits on their CP is required to confidently assert whether their AGN has switched off. As such, these sources must retain their remnant candidate classification.  \\

A low core-prominence criterion is indeed necessary to classify recently switched off remnants. However, we find that the three remnant candidates with the weakest constraints on their CP upper limits are also those without hotspots, two of which have ultra-steep spectra, e.g. see Sect.~\ref{sec:remnants_without_hotspots}. While the sample size is rather small, this observation would suggest that aged remnants may be preferentially deselected if only a low ($\lesssim10^{-4}$) CP criterion is used. This observation echoes the results of \citep{2017A&A...606A..98B}, who show that none of their ultra-steep spectrum remnants are selected by low CP (e.g.~$<0.005$), and, only 3/10 morphologically-selected remnants are selected by low CP.


\begin{figure}
    \centering
    \includegraphics[width=\linewidth]{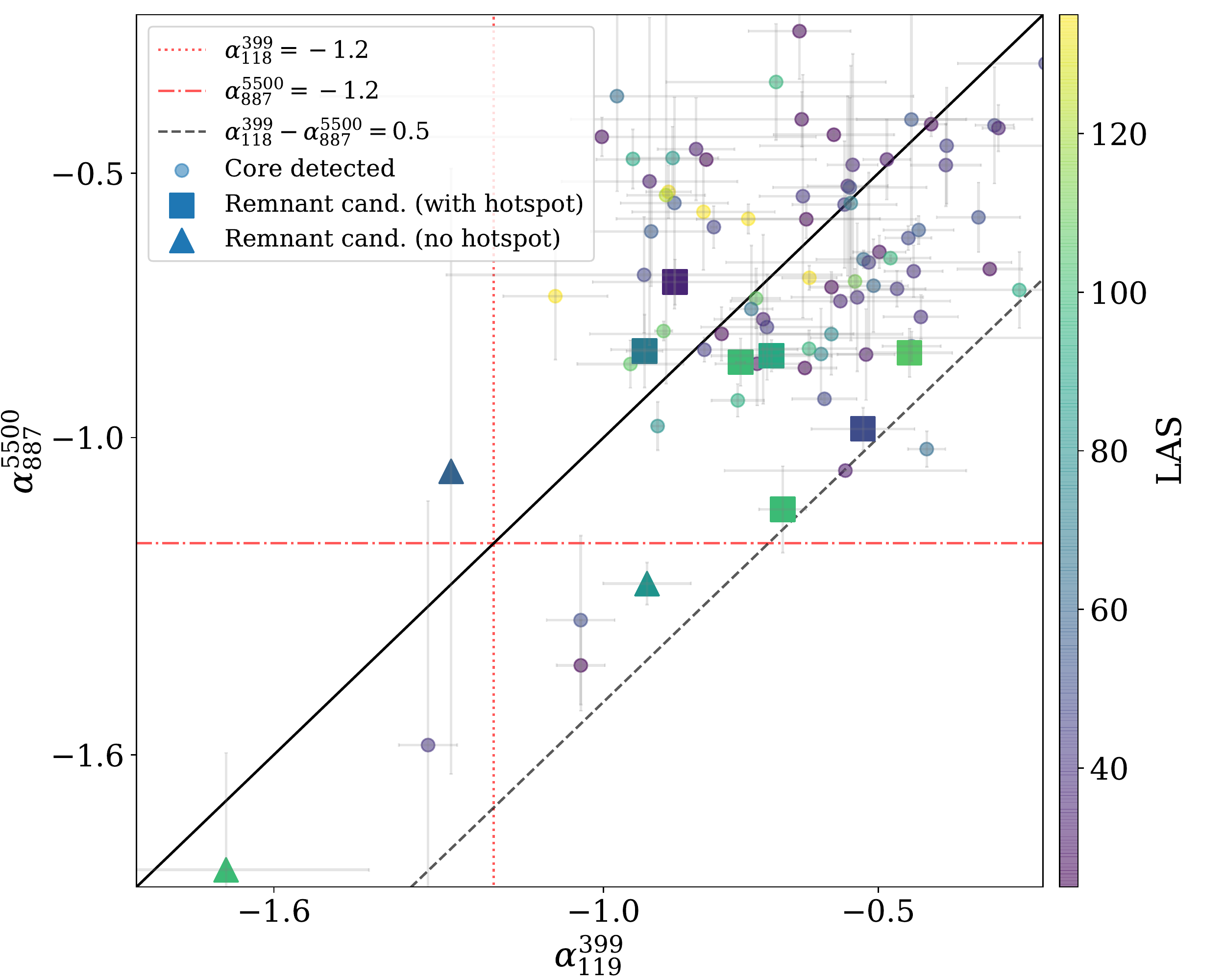}
    \caption{The high-frequency spectral index $\alpha_{887}^{5500}$ is plotted against the low-frequency spectral index $\alpha_{119}^{399}$. A third color-bar axis is over-plotted to show the largest angular size in arc-seconds. Solid black line represents a constant spectral index across both frequency ranges. Dashed black line represents a spectral curvature of $\mathrm{SPC}=0.5$. The red dotted and dot-dashed lines represent a $\alpha=-1.2$ spectral index across the low- and high- frequency range, respectively.}
    \label{fig:alpha_alpha}
\end{figure}
\subsubsection{Spectral index distribution}
\label{sec:alpha}
\begin{table*}
    \centering
    \caption{Spectral index statistics calculated based on data represented in Figure~\ref{fig:alpha_alpha}. The median and mean spectral index, indicated by $_{\mathrm{med}}$ and $_{\mathrm{mean}}$ subscripts, are presented for the low $\alpha_{119}^{399}$ and high $\alpha_{887}^{5500}$ frequency ranges. $f_{\mathrm{US,\:low}}$ and $f_{\mathrm{US,\:high}}$ represent the low- and high-frequency ultra-steep fractions, respectively.\\
    $^\dagger$ A range is given here, as it is unclear whether MIDAS~J225608$-$341858 (Fig.~\ref{fig:MIDAS_J225608-341858}) is ultra-steep at high frequencies.}
    \label{tab:alpha}
    \begin{tabular}{lcccccc}
    \hline 
    
    \hline
    Sample &$\alpha_{\mathrm{119, \: med}}^{399}$&$\alpha_{\mathrm{119, \:mean}}^{399}$&$\alpha_{\mathrm{887, \:med}}^{5500}$&$\alpha_{\mathrm{887, \:mean}}^{5500}$ &$f_{\mathrm{US,\:low}}$ &$f_{\mathrm{US,\:high}}$\\
    \hline
    \hline
    Core-detected & -0.60 & -0.63 & -0.66 & -0.67 & 0/94&3/94 \\
    Remnant cand. (with hotspot) & -0.69 & -0.69 & -0.84 & -0.88 & 0/7& 1/7\\
    Remnant cand. (without hotspot) & -1.27 & -1.29 & -1.27 & -1.38 & 1/3& $(2-3)^\dagger/3$\\
    \hline
    \hline
    \end{tabular}
\end{table*}
The integrated spectral properties of our sample are explored over two frequency ranges. The integrated flux densities at 119, 154, 186, 216 and 399\,MHz are used to develop a low-frequency spectral index, $\alpha_{119}^{399}$. To compute correct fitting uncertainties, we fit power-law models to the data in linear space. A high-frequency spectral index, $\alpha_{887}^{5500}$, is computed using $\alpha=\frac{\mathrm{log_{10}}(S_{887}/S_{5500})}{\mathrm{log_{10}}(887/5500)}$, and an associated uncertainty $\Delta\alpha=\frac{1}{\mathrm{ln}(887/5500)}\sqrt{\big(\frac{\Delta S_{887}}{S_{887}}\big)^2 + \big(\frac{\Delta S_{5500}}{S_{5500}}\big)^2}$. We populate our results onto an $\alpha-\alpha$ plot, e.g. Figure~\ref{fig:alpha_alpha}, and summarize our results in Table~\ref{tab:alpha}.

Despite the access to high frequencies ($\nu$ = 5.5 GHz), we find that the selected remnant candidates with hotspots show similar spectral properties to radio galaxies with an active radio core. At low frequencies, remnant candidates with hotspots display spectral indices that are consistent with continuous injection. The high frequency spectral index does appear to be steeper than for the bulk of remnant candidates with hotspots, however, this is also observed for core-detected radio sources and simply reflects the preferential ageing of higher-energy electrons. No remnant candidates with hotspots display an ultra-steep low-frequency spectral index, and only one such source (e.g. Sect.~\ref{sec:MIDAS_J225337-344745}) demonstrates a high-frequency ultra-steep spectral index. Regarding these remnant candidates with non ultra-steep spectra, their position on the $\alpha-\alpha$ plot can be explained if these are young, recently switched off remnants. However, their spectral properties can just as easily be explained if these are active radio galaxies in which the radio core is below the GLASS detection limit. We can not rule out either of these possibilities based on their spectra alone, and as such they must remain as remnant candidates. We note that \citet{2018MNRAS.475.4557M} also report on a large overlap in the observed spectral index (150\,MHz--1.4\,GHz) between their active and candidate remnant radio galaxies. Only one of their remnant candidates display hotspots at 6\,GHz with the Very Large Array telescope, meaning it is not necessarily only the remnant candidates with hotspots which display similar spectral indices as active radio galaxies. A great example of this is the remnant radio galaxy Blob~1 identified by \citet{2016A&A...585A..29B}. \\

We also find a small fraction (3/94) of core-detected radio sources which demonstrate an ultra-steep spectral index. The angular size of these sources is well below the largest angular scale that GLASS can recover at 5.5\,GHz, suggesting the curvature is genuine. This tells us that ultra-steep selection will not only select radio galaxies in which the AGN has switched off, however it is interesting to note these sources also do not display GLASS hotspots. It is possible these sources represent restarted radio galaxies in which the `core' represents a newly-restarted jet. \\

Unsurprisingly, we find that the remnant candidates without hotspots also display steeper spectra than those with hotspots. The absence of compact features in their morphologies implies a lack of recent jet activity, and is supported by their ultra-steep spectra which implies significant spectral ageing within the lobes. Our low fraction of ultra-steep spectrum remnant candidates echoes the result of \citet{2016A&A...585A..29B} who discuss the bias of preferentially selecting aged remnant radio galaxies via their ultra-steep spectra.\\


\begin{table*}
    \centering
    \caption{Derived distribution averages from Sect.~\ref{sec:restframe}. The number of radio sources included in each category are denoted by $N$. The redshift, $z$, radio power, $P$, and largest linear size (LLS) are presented. The subscripts med and mean refer to the median and mean values. In the upper half of the table, we consider the entire sample of 104 radio sources. In the lower half, we consider only those with spectroscopic redshifts}.\\$^\dagger$ Including the 14 core-detected radio galaxies with $z\geq1$ lower limits.
    \label{tab:rest_frame_stats}
    \begin{tabular}{lccccccc}
    \hline
    Sample &$N$&$z_{\mathrm{med}}$&$z_{\mathrm{mean}}$ &$P_{\mathrm{med}}$ &$P_{\mathrm{mean}}$ &$\mathrm{LLS}_{\mathrm{med}}$ & $\mathrm{LLS}_{\mathrm{mean}}$\\
    & & & & log$_{10}(\mathrm{W\,Hz^{-1}})$ & log$_{10}(\mathrm{W\,Hz^{-1}}$) &kpc &kpc \\
    \hline
    \textit{Full sample.}\\
    \hline
    Core-detected & 80& 0.519 & 0.59 & 25.2 & 25.2 & 277 & 379 \\
    Core-detected$^\dagger$ &94& 0.549 & 0.651 & 25.3 & 25.3 & 294 & 388 \\
    Remnant candidates &10& 0.504 & 0.619 & 25.5 & 25.8 & 435 & 512 \\
    \hline
    \textit{Spectroscopic redshifts.} \\
    \hline
    Core-detected &25& 0.303 & 0.290 & 25.1 & 25.2 & 171 & 322 \\
    Remnant candidates  &3& 0.312 & 0.279 & 25.2 & 25.1 & 351 & 299 \\
    \hline
    \end{tabular}
\end{table*}
\subsubsection{Power--size distribution}
\label{sec:restframe}
\begin{figure}
    \centering
    \includegraphics[width=\linewidth]{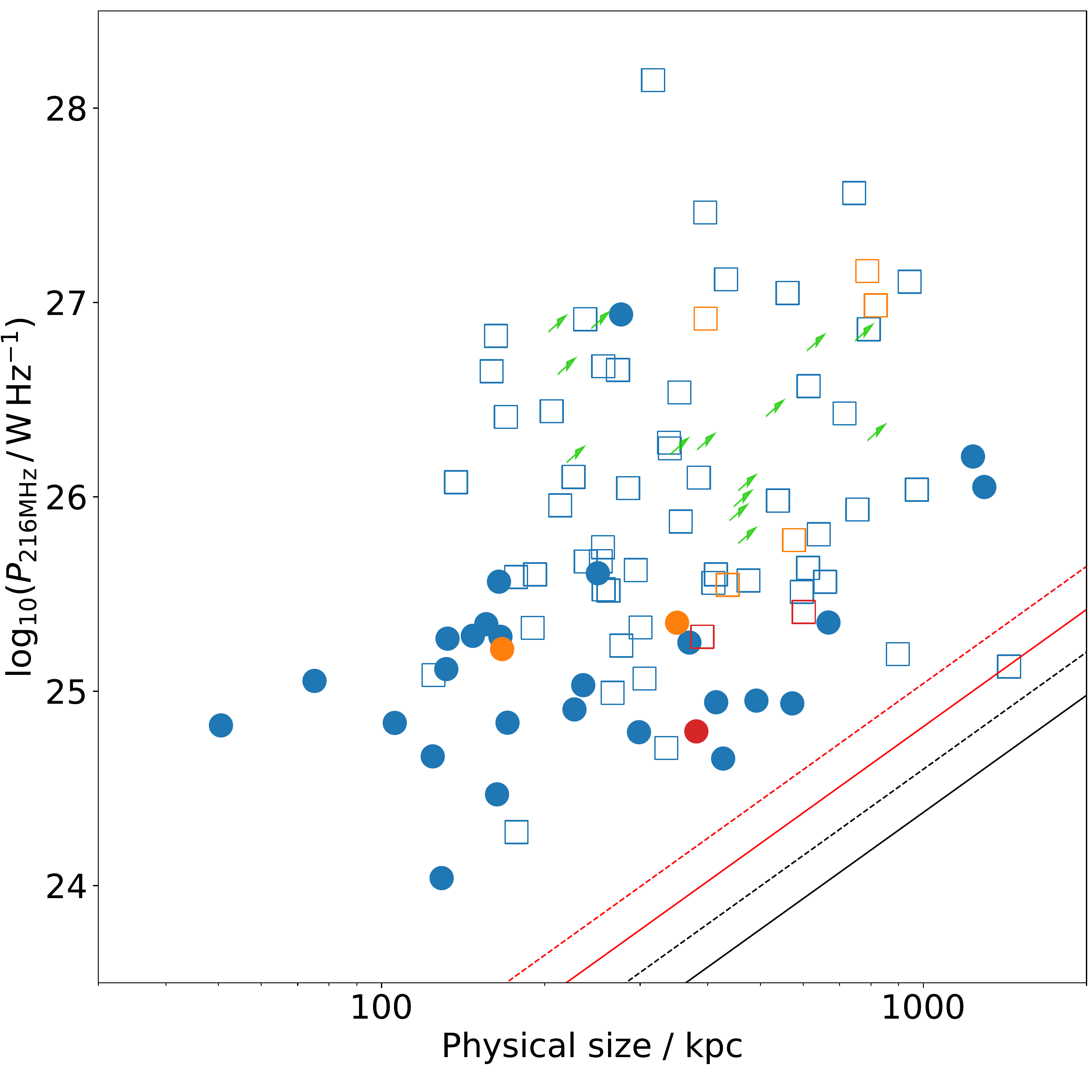}
    \caption{216\,MHz radio power against the largest linear size. Core-detected radio sources (blue markers), remnant candidates with hotspots (orange markers) and remnant candidates without hotspots (red markers) are displayed. Circular and square markers are used to denote spectroscopic and photometric redshifts, respectively. Lower limits on the 14 radio sources without host identifications are denoted by green arrows. Plotted also are the largest linear sizes that would result in a 5$\sigma$ detection at 216\,MHz at $z=0.3$ (black) and $z=1$ (red). Limits are calculated assuming a uniform brightness ellipse, and a lobe axis ratio of 2.5 (solid line) and 1.5 (dashed line). Aged remnants often display low axis ratios, e.g. MIDAS~J225522$-$341807 (Sect.~\ref{sec:MIDAS_J225522-341807}).}
    \label{fig:pvsd}
\end{figure}

We investigate the sample distribution in redshift, total radio power, and largest linear size, given the host galaxy identifications. We stress that many of the selected remnant candidates have uncertain host galaxy associations, presenting a major challenge in analysing their rest-frame properties. An additional uncertainty comes from the photometric redshifts, which make up $60/104$ (57\%) of the sample. In addition, 14 active radio galaxies do not have an optical identification, meaning photometric redshift estimates can not be attained. If we assume their host galaxies are at least $10^{10.6}$M$_{\odot}$, e.g. the lowest stellar mass reported by \citet{2005MNRAS.362...25B} to host a radio-loud AGN, we can apply the $K-z$ relation (e.g. see \citealt{1984JApA....5..349L}, \citealt{2004A&A...415..931R}) to estimate the lowest redshift for which a $10^{10.6}$M$_{\odot}$ galaxy will be undetected below the VIKING $K_s$-band AB magnitude limit (Sect.~\ref{sec:vik}). This suggests that the host galaxies of the 14 unidentified radio galaxies must be above $z = 1$. The caveat here is that \citet{2005MNRAS.362...25B} investigate local ($z\leq0.3$) radio-loud AGN samples, whereas here we are assuming higher redshift. Figure~1 of \citet{2017A&A...602A...6S} shows a hint of a decline in stellar mass with redshift, however, our assumption of the minimum stellar mass seems valid up to $z\sim1$. For each radio source, we calculate the total radio power following: $P= 4\pi S D_L^2(1+z)^{-\alpha - 1}$, where $D_L$ is the luminosity distance. For the 14 radio galaxies without optical identifications, the radio power is calculated assuming $z\geq1$. Our results are presented in Table~\ref{tab:rest_frame_stats} and Figure~\ref{fig:pvsd}.\\

Assuming radio lobes continue expanding once the jets switch off, which at least appears to be the case for FR-II radio galaxies as shown by \citet{2017MNRAS.471..891G}, an expectation of this is for remnants to display larger physical sizes with respect to their active radio galaxy progenitors. Interestingly, our results suggest that the largest linear sizes of core-detected radio galaxies and remnant candidates are similar. This may be explained by the observational bias against remnant radio galaxies of large linear size; such sources will preferentially fall below a fixed detection limit due to their lower surface brightness profiles. Since the linear size and age of a radio galaxy are correlated, for a fixed jet power and environment, it is not unreasonable to suggest that these `missing' remnants also correspond to older remnant radio galaxies, which in turn would imply our sample predominately comprises of young remnants (see also \citealt{2018MNRAS.475.4557M}). If so, this result is consistent with the low fraction of ultra-steep remnants discussed in in Sect.~\ref{sec:alpha}. We do however caution this analysis since seven remnant candidates have ambiguous host galaxy associations, as well as the uncertainties surrounding the photometric redshift estimates.\\

We instead consider the sample of 28 spectroscopically-confirmed radio galaxies to see whether we can draw the same conclusions as above. Within this `spectroscopic sample', we can be confident not only of the sample redshifts, but also of the three remnant candidates (e.g. see Sect.~\ref{sec:MIDAS_J225607-343212}, \ref{sec:MIDAS_J225337-344745}, and \ref{sec:MIDAS_J230104-334939}) which have unambiguous host galaxy associations, meaning we can be confident of their positions on the power-size diagram. As demonstrated in Figure~\ref{fig:pvsd}, the absence of remnant radio galaxies of large linear size becomes quite clear in the `spectroscopic sample', and is consistent with the previously-discussed conclusions. Our limiting factor here is the small sample size, however this will be addressed in future work where we can expect a factor $\sim6$ increase in the sample size of radio galaxies in GAMA~23. The additional benefit here will be the GAMA~23 group catalogue, which provides group/cluster associations for galaxies with spectroscopic redshifts, as well as virial estimates for the group mass. This will allow us to begin decomposing the degeneracy between radio power, linear size and environment.\\

\begin{table*}
\centering
\caption{Remnant fractions constrained by previous authors. Each column, in ascending order, represents the cited study, the sky coverage over which the sample is compiled, the flux limit across the sample (or the faintest source in the sample), the frequency at which the flux cut is made, the angular size cut of the sample, the number of radio galaxies within the sample, and the resulting remnant fraction.\\ \textbf{References.} (1) \citet{2012ApJS..199...27S}, (2) \citet{2017A&A...606A..98B} and \citet{2020arXiv200409118J}, (3) \citet{2018MNRAS.475.4557M}, (4)~This~work.}
\label{tab:remfracpapers}
\begin{tabular}{lcccccc}
\hline
Ref. & Sky coverage & Flux limit & Sample frequency &$\theta_{\mathrm{cut}}$ & Sample size & $f_\mathrm{rem}$\\
& (deg$^2$) & (mJy) & (MHz)& ($''$) & & \\
\hline
\hline
1 & 7.52 & 1 & 1400 &30 & 119 & <~4\%\\
2 & 35 & 40 & 150  & 60 & 158 & <~11\% \\
3 & 140 & 80 & 150 & 60 & 127& <~9\% \\
4 & 8.31 & 10 & 216 & 25 & 104 & $4\lesssim f_{\mathrm{rem}} \lesssim 10\%$ \\
\hline
\end{tabular}
\end{table*}

\subsection{Constraining a remnant fraction}
\label{sec:remfrac}

\subsubsection{Remnant fraction upper limit}
The fraction of remnant radio galaxy candidates identified in this work provides an upper limit to the genuine fraction of remnant radio galaxies, $f_\mathrm{rem}$, present within this sample. Of 104 radio galaxies, 10 are identified as remnant radio galaxy candidates, resulting in a $f_\mathrm{rem}\approx10\%$ upper limit on the remnant fraction. \citet{2012ApJS..199...27S}, \citet{2017A&A...606A..98B} and \citet{2018MNRAS.475.4557M} each constrain a remnant fraction from radio observation and their results are presented in Table~\ref{tab:remfracpapers}. At face value, the remnant fraction obtained in this work is consistent with that of \citet{2017A&A...606A..98B} and \citet{2018MNRAS.475.4557M}, and shows a considerable increase over the fraction constrained by \citet{2012ApJS..199...27S}. The apparent inconsistency with  \citet{2012ApJS..199...27S} may very well be a result of their selection. Their sample was selected at 1.4\,GHz, where ultra-steep remnants may potentially be missed, and they also excluded sources without a radio core but hotspots still present within the lobes. It is interesting to note that despite the difference in the flux limit and angular size cut compared to the samples complied by \citet{2018MNRAS.475.4557M}, \citet{2017A&A...606A..98B}, and \citet{2020arXiv200409118J}, the upper limits on the remnant fraction appear consistent. \citet{2020MNRAS.tmp.1303S} show that the remnant fraction predicted by constant-age models, e.g. those in which the jets are active for a constant duration, is highly sensitive to observable constraints, e.g. the flux and angular size limit. On the other hand the remnant fraction predicted by power-law age models, e.g. those in which the duration of the active phase is power-law distributed, shows little dependence on observable parameters. It is therefore possible that the similarity in remnant fractions implies a preference towards power-law age models that describe AGN jet activity, although this needs to be pursued in more detailed future work.

\subsubsection{Remnant candidates without hotspots}
\label{sec:remfrac_revised}
As presented in Sect.~\ref{sec:results}, the lobes of seven remnant candidates display compact emitting regions in GLASS, potentially indicating a hotspot formed by an active jet. As discussed in Sections~\ref{sec:cp}~and~\ref{sec:alpha}, these sources could be interpreted either as recently switched off remnants, or, active radio galaxies with unidentified radio cores. We thus propose a lower limit to the remnant fraction, by considering the limiting case where each remnant candidate with a hotspot is an active radio galaxy. This would suggest a $f_{\mathrm{rem}}=4/104$ ($\approx4\%$) lower limit on the remnant fraction, consistent with the value reported by \citet{2012ApJS..199...27S}. As discussed in Sect.~\ref{sec:evolutionary}, an appreciable fraction of genuine remnants with hotspots can potentially be expected.

\begin{figure*}
    \begin{subfigure}{0.5\linewidth}
    \centering\includegraphics[width=\linewidth]{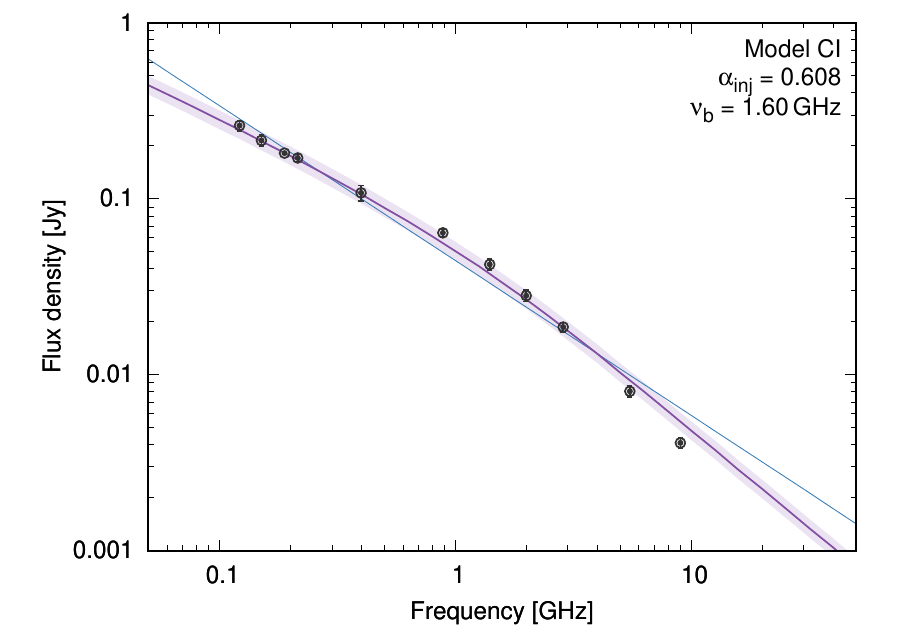}
    \caption{Continuous injection model (CI)}
    \label{fig:ci}
    \end{subfigure}\hfill
    \centering
    \begin{subfigure}{0.5\linewidth}
    \centering\includegraphics[width=\linewidth]{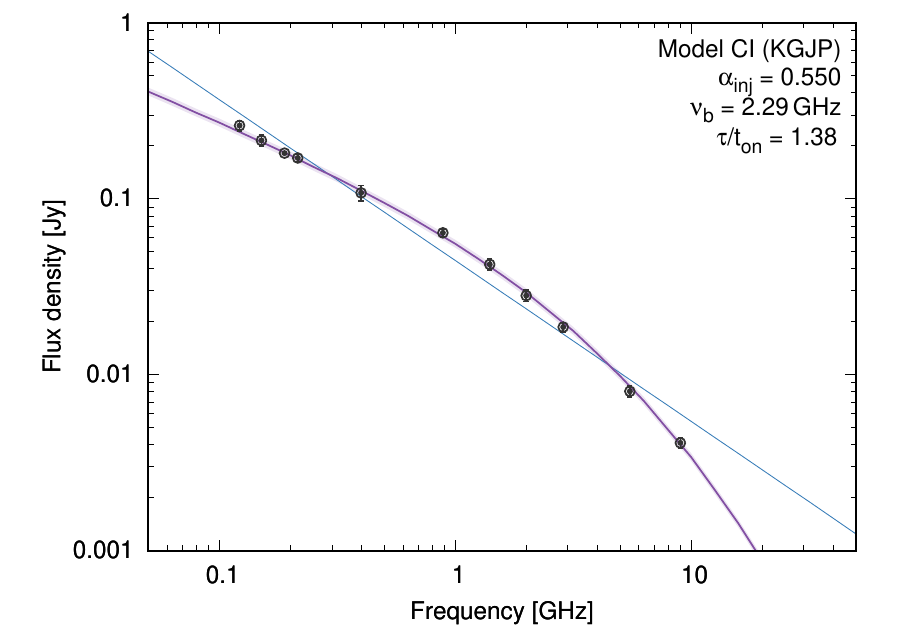}
    \caption{Continuous injection with `off' component model (CI off)}
    \label{fig:cioff}
    \end{subfigure}
    \caption{Modelled integrated spectrum of MIDAS~J225337$-$344745. Figure~\ref{fig:ci} models the spectrum assuming a continuous injection model (CI). Figure~\ref{fig:cioff} models the spectrum assuming a continuous injection model with an `off' component, encoding a jet switch-off (CI off). In each model, a 2$\sigma$ uncertainty envelope is represented by the violet shaded region. As discussed in Sect.~\ref{sec:evolutionary}, the model uncertainties take into account only the uncertainties on the flux density measurements, and do not reflect the underlying uncertainties due to an inhomogeneous magnetic field. For reference, a best-fit to the data using single power-law model is represented by a blue line.}
\label{fig:models}
\end{figure*}

\begin{figure}
    \centering
    \includegraphics[width=\linewidth]{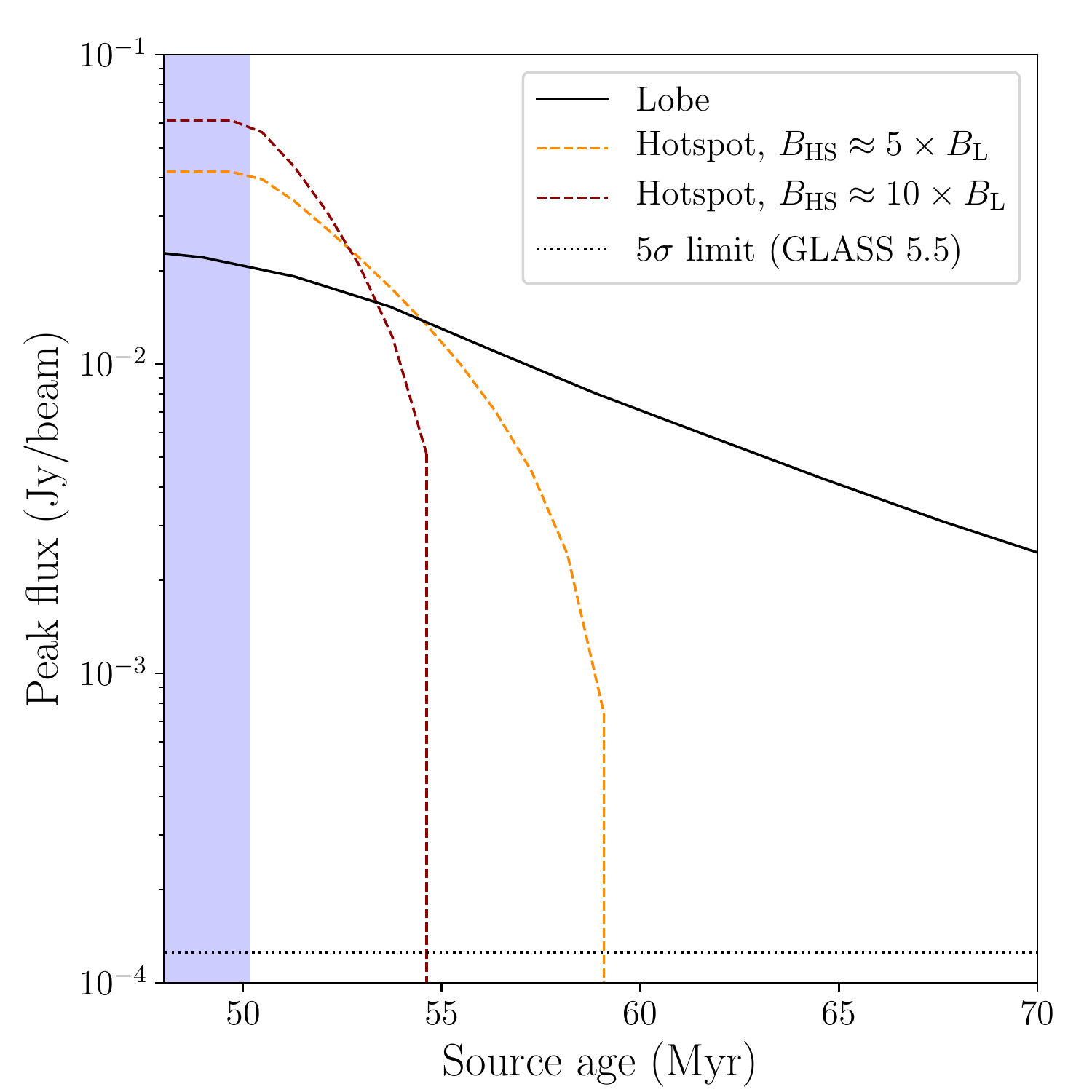}
    \caption{A `MIDAS~J225337$-$344745' type remnant radio galaxy is modelled by assuming a jet power $Q=10^{38.1}$\,W, an injection energy index of $s=2.1$, an equipartition factor of $B/B_{\mathrm{eq}}=0.22$, and a total source age of 71\,Myr of which 50\,Myr is spent in an active phase, and a further 21\,Myr is spent as a remnant. The shaded blue bar corresponds to the time during which the source is active, after which the jets are switched off and the hotspots/lobes begin to fade. The evolution of the synchrotron emission from the lobes (solid black tracks) and the hotspots (dashed tracks) are shown as a function of the total source age. The assumption that the hotspot magnetic field strength is a factor five greater than the lobes (colored~in~orange) comes from Cygnus~A, however we also assume a factor ten increase in the hotspot magnetic field strength (colored~in~red) to consider shorter fading timescales. We explore this in terms of the peak flux density, as this ultimately decides whether the emitting regions are detected in observations. The vertical drop in the flux density tracks reflects the depletion of electrons capable of producing emission at 5.5\,GHz. As expected, the synchrotron emission evolves faster in the hotspot, however, their fading timescale is non-negligible in comparison to that of the lobes.}

    \label{fig:hs_lb}
\end{figure}
\begin{table}
\centering
\caption{Summarized properties of MIDAS J225337$-$344745 spectral modelling (Section \ref{sec:evolutionary}). A reduced chi-squared ($\chi^2_{\mathrm{red}}$) is provided to assess the quality of fit. The injection index $\alpha_{\mathrm{inj}}$, observed-frame break frequency $\nu_b$ and quiescent fraction $T$ are presented for the fitted continuous injection (CI) and continuous injection-off CI-off models. We quote a $\Delta$BIC calculated between the two models.}
\label{tab:ci_models}
\begin{tabular}{lccccc}
\hline
Model & $\chi^2_{\mathrm{red}}$ &$\alpha_{\mathrm{inj}}$ & $\nu_b$  & $T$ & $\Delta$BIC
\\
fitted & & & GHz &\\
\hline
\footnotesize{CI} & \footnotesize{3.43}& \footnotesize{-0.608} & \footnotesize{1.60$\pm$0.08} & $-$ &\multirow{2}{*}{\footnotesize{2.1}}\\
\footnotesize{CI-off} & \footnotesize{0.64}& \footnotesize{-0.55} & \footnotesize{2.2$\pm$0.09} & \footnotesize{0.28$\pm$0.004}\\

\hline
\end{tabular}
\end{table}

\subsection{Evolutionary history of MIDAS~J225337-344745}
\label{sec:evolutionary}

A particularly interesting source to examine is MIDAS~J225337$-$344745, which as mentioned in Sect.~\ref{sec:MIDAS_J225337-344745}, demonstrates two seemingly contradictory features: the higher-frequency ultra-steep ($\alpha<-1.52$) spectrum of the lobes, suggesting energy losses consistent with an aged remnant radio galaxy, and the compact 5.5\,GHz emitting regions in GLASS, which may in turn suggest current or recent energy injection. It is unclear whether these are genuine hotspots; the north-eastern lobe is detected just above the noise level and thus demonstrates many `hotspot-like' features. A singular, bright `hotspot' is evident in the south-western lobe, however, the emitting region is clearly resolved at a physical resolution of $7\times14$\,kpc (the physical resolution of GLASS~5.5 at $z=0.213$), and so it is unclear whether this `hotspot' is formed by an active jet, or, whether it is the expanding hotspot of a previously-active jet. It is also possible that the compact emitting region is a combination of the lobe and hotspot. Fortunately, the radio continuum spectrum appears to preserve the original spectrum at low frequencies, while also encoding the energy losses evident at higher frequencies. This allows us to model and parameterise the spectrum in terms of physically meaningful parameters, allowing us to probe the energetics along with the dynamics. Our ultimate objective here is to determine whether the compact emitting regions can be expected within a remnant radio galaxy. Our analysis is laid out as follows. 


We model the radio source using the \textit{Radio AGN in Semi-analytic Environments} (RAiSE) code which considers, amongst other things: (i) the evolution of the magnetic field strength; (ii) the adiabatic losses contributing to the radio luminosity evolution of the lobes; and (iii) the Rayleigh-Taylor mixing of the plasma contained within the remnant lobes. These processes are described by the dynamical \citep{2015ApJ...806...59T} and synchrotron emissivity \citep{2018MNRAS.474.3361T} models for lobes FR-II radio lobes, which follow the continuous injection model \citep[CI model;][]{1962SvA.....6..317K,1970ranp.book.....P,1973A&A....26..423J}. The remnant spectrum follows the continuous injection `off' model \citep[CI-off;][]{1994A&A...285...27K} alternatively known as the KGJP model, which \citet{2018MNRAS.476.2522T} parameterise with two break frequencies. A jet power \textit{Q}$=10^{38.1}$\,W, magnetic field strength $B=1.08$\,nT, and total source age $\tau=71$\,Myr are constrained by the method of \citet{2018MNRAS.474.3361T}. The synchrotron spectrum is modelled by the method of \citet{2018MNRAS.476.2522T} which, provided sufficient spectral sampling, allows the injection index $\alpha_{\mathrm{inj}}$, break frequency $\nu_b$ and quiescent fraction\footnote{The quiescent fraction is defined as $T=t_{\mathrm{off}}/\tau$, where $t_{\mathrm{off}}$ is the duration of the remnant phase, and $\tau$ is the total age of the radio source.} $T$ to be uniquely constrained. Our results are shown in Figure~\ref{fig:models}. Both models constrain an injection index that is consistent within their typically observed range of $-1.0<\alpha_{\mathrm{inj}}< -0.5$ (see Table~\ref{tab:ci_models}). The discrepancy between the modelling becomes evident at the highest frequencies, for which the CI model begins to over-predict the integrated flux density. On the other hand, the CI-off model is able to comfortably reproduce the observed spectral curvature. We calculate a reduced~chi$^2$ ($\chi^2_{\mathrm{red}}$) for each model, and a $\Delta$\,BIC$=2.1$ (by BIC$_{\mathrm{CI}}$ $-$ BIC$_{\mathrm{CI-off}}$) which demonstrates a preference towards the CI-off model (see Table~\ref{tab:ci_models}). Although the distinction between the two models is ultimately driven by the 9\,GHz measurement, CI-off provides a better model of the observed spectrum suggesting these are remnant lobes. Modelling the spectrum as a remnant suggests the source spends approximately 21\,Myr in a remnant phase.

\citet{2018MNRAS.474.3361T} showed that the CI model provides a statistically significant fit to the broad frequency radio spectra of over 86 per cent of FR-IIs in the \citet{2008MNRAS.390..595M} sample of 3C sources (typically 12 measurements between 0.01 and 10 GHz from \citealt{1980MNRAS.190..903L}; but see \citealt{2017MNRAS.466.2888H}). Further, despite the non-physical CI model assumption of time-invariant magnetic fields, \citet{2018MNRAS.474.3361T} find that CI model is an excellent fit to the simulated spectra of lobed FR-II which do consider magnetic field evolution, and are able to recover the source dynamical age.

As a final sanity check on the duration of the remnant phase fitted above (by the integrated CI-off model), we constrain the age of the most recently accelerated electron populations which we assume are located near the `hotspots'. We convolve the radio observations at 399\,MHz, 887\,MHz and 5.5\,GHz images to a common 11$''$ circular resolution and measure the integrated flux density within an aperture centered at the southern `hotspot'. We fit the spectrum with a Tribble JP model assuming the magnetic field derived previously in our RAiSE modelling. We arrive at a remnant age of $t_{\mathrm{off}}=13_{-4}^{+8}$\,Myr, consistent with the $t_{\mathrm{off}}\approx21$\,Myr derived from the integrated CI-off model. We stress that the observations are not properly \textit{(u,v)} matched, and thus are not necessarily seeing the same radio emission; e.g. at 5.5\,GHz, the flux density is likely underestimated due to resolving out the extended radio emission. We leave a more detailed analysis of this source to future work.

Next, we model the hotspot to better understand its typical fading timescale. 
We model a `MIDAS~J225337$-$344745 like' radio galaxy, meaning we adopt the same values for the jet power, active age, lobe axis ratio, energy injection index and equipartition factor. By the method of \citet{2018MNRAS.473.4179T} we forward model the 5.5\,GHz lobe spatial emissivity (without the hotspot) and convolve the output map with the GLASS~5.5 synthesized beam. We consider the GLASS~5.5 beam with the largest integrated flux (e.g. the maximum peak flux density), as this ultimately determines whether the youngest emitting regions of the lobes are detected. The hotspot is modelled assuming a JP spectrum \citep{1973A&A....26..423J} for the same initial properties as the lobe, but for an increased magnetic field strength. Cygnus~A displays a factor five increase in the hotspot magnetic field strength, in comparison to that of the lobes \citep{1996A&ARv...7....1C}. We make the same assumption for MIDAS~J225337$-$344745, however also consider a case where the hotspot magnetic field strength is a factor 10 greater than the lobes. We also assume the same ratio of hotspot to lobe volume as that measured in Cygnus~A; this only shifts the spectrum along the log-frequency log-luminosity plane and does not influence the spectral shape (i.e. the slope). Our results are presented in Figure~\ref{fig:hs_lb}. As expected, the hotspot fades rapidly once injection switches off. Bright radio emission from the hotspot is driven by the freshly injected electrons which have a short lifetime at high frequencies due to the strong magnetic field. Despite the hotspot fading rapidly, its fading timescale is non-negligible when compared to the characteristic active and remnant ages. It is not unreasonable to suggest that recently switched off remnants may still display hotspots, and that some of the selected remnant candidates with hotspots (Sect.~\ref{sec:remfrac_revised}) may indeed be such cases. The main result of Figure~\ref{fig:hs_lb} is that the hotspots of a recently switched off jet should \textit{not} be visible in MIDAS~J225337$-$344745, assuming a $t_{\mathrm{off}}\approx21$\,Myr remnant phase. At the implied age of the remnant, the modelled emitting regions of the lobes range between 100--150\,kpc in size. This is consistent with the observed emitting regions in GLASS, which demonstrate a projected linear size of $\sim96$\,kpc. This indicates that the bright features in GLASS are entirely consistent with the youngest plasma regions, and are not necessarily the hotspots of an active jet. 

As such, the interpretation of this source is challenging for the following reasons. Modelling the spectrum suggests these are the lobes of a remnant radio galaxy. While hotspots will remain visible for a non-negligible period of time after injection is switched off, they should not be visible in this source assuming the remnant age is correct. The two pieces of evidence are consistent with one another if the bright GLASS features are just the youngest emitting regions of the lobes. If we alternatively assume MIDAS~J225337$-$344745 is an active source, the RAiSE modelling of \citet{2018MNRAS.474.3361T} finds a comparable age ($\tau = 100\rm\: Myr$) to the remnant case but a substantially lower jet power of $Q=10^{36.7}\rm\: W$ is required to match the relatively low observed flux density. This jet power is typically associated with an FR-I morphology for the known host galaxy mass \citep[e.g.][]{1994AAS...184.1702L,2009AN....330..184B}, potentially resulting in a broad flare point rather than compact hotspots at the end of a jet. This morphology may explain the large spatial extent of the GLASS 5.5\:GHz observations, however does not resolve the discrepant spectral curvature. 

Finally, the result presented in Figure~\ref{fig:hs_lb} makes a rather interesting prediction for the observed properties of genuine remnants. As discussed in Sect.~\ref{sec:remfrac}, the upper limit on the remnant fraction is $f_{\mathrm{rem}}\approx10\%$. For a \textit{characteristic} active lifetime of $\sim100$\,Myr, our observed remnant fraction would thus suggest an observable remnant phase lasting $\sim10$\,Myr. The modelling of a `MIDAS~J225337$-$344745 like' radio galaxy shows that the hotspot can persist for at least several Myr before fading out completely. As such, we can expect an appreciable fraction of remnants to still display hotspots. 
\section{Conclusions}
Within a sub-region of 8.31\,deg$^2$ of the GAMA~23 field, we have compiled a sample consisting of 104 extended, low-frequency selected (216\,MHz) radio galaxies. Using the 5.5 and 9.5-GHz GLASS survey, we have adopted the `absent radio core' criterion to search for remnant radio galaxy candidates. Our conclusions are summarized as follows:
\label{sec:conclusion}
\begin{itemize}
    \item We identify 10 new remnant radio galaxy candidates, thereby constraining an $f_{\mathrm{rem}}~\leq~10\%$ upper limit on the fraction of radio galaxies with quiescent AGN. Our upper limit is consistent with that proposed by previous authors, and suggests that remnants must have a short observable lifetime. 
    \item Seven remnant candidates show compact emitting regions in GLASS, an observation that can only be explained if the jets have recently switched off. A much smaller fraction (3/10) show relaxed, hotspot-less lobes, and only one displays an ultra-steep spectrum across the entire frequency range. This implies remnants are detected soon after switching off, suggesting a rapid fading during the remnant phase.
    \item The small fraction of ultra-steep ($\alpha<-1.2$) remnants is likely a result of the oldest remnant lobes escaping detection due to their expansion.
    \item At present, the upper limits placed on the remnant core prominence are too weak to confidently rule out the presence of AGN activity. Those with compact hotspots and a non ultra-steep spectrum must therefore retain their remnant candidate classification. Considering the limiting case in which \textit{all} these are active radio galaxies, we would expect a $f_{\mathrm{rem}}\approx4\%$ remnant fraction.  
    \item MIDAS~J225337$-$344745 represents an interesting object for future study. Modelling the integrated lobe spectrum shows consistency with a remnant radio galaxy. We find that although the hotspot has a non-negligible fading timescale, we do not expect to see hotspots in this source.
    \item By modelling the synchrotron spectrum arising from a `MIDAS~J225337$-$344745-like' hotspot, we show that the hotspot can persist for $5-10$\,Myr at 5.5\,GHz after the jets switch off. This would imply that the presence of a hotspot in radio maps may not necessarily reflect an active jet, and by extension we can expect an appreciable fraction of genuine remnants to still display hotspots.
    
\end{itemize}
\section{Acknowledgements}
BQ acknowledges a Doctoral Scholarship and an Australian Government Research Training Programme scholarship administered through Curtin University of Western Australia. NHW is supported by an Australian Research Council Future Fellowship (project number FT190100231) funded by the Australian Government. This scientific work makes use of the Murchison Radio-astronomy Observatory, operated by CSIRO. We acknowledge the Wajarri Yamatji people as the traditional owners of the Observatory site. Support for the operation of the MWA is provided by the Australian Government (NCRIS), under a contract to Curtin University administered by Astronomy Australia Limited. We acknowledge the Pawsey Supercomputing Centre which is supported by the Western Australian and Australian Governments. The Australian SKA Pathfinder is part of the Australia Telescope National Facility which is managed by CSIRO. Operation of ASKAP is funded by the Australian Government with support from the National Collaborative Research Infrastructure Strategy. The Australia Telescope Compact Array is part of the Australia Telescope National Facility which is funded by the Australian Government for operation as a National Facility managed by CSIRO. We thank the staff of the GMRT that made these observations possible. GMRT is run by the National Centre for Radio Astrophysics of the Tata Institute of Fundamental Research. CHIC acknowledges the support of the Department of Atomic Energy, Government of India, under the project  12-R\&D-TFR-5.02-0700. SW acknowledges the financial assistance of the South African Radio Astronomy Observatory (SARAO) towards this research is hereby acknowledged (\url{www.ska.ac.za}). IP acknowledges support from INAF under the SKA/CTA PRIN "FORECaST" and the PRIN MAIN STEAM "SAuROS" projects. The National Radio Astronomy Observatory is a facility of the National Science Foundation operated under cooperative agreement by Associated Universities, Inc. H.A. benefited from grant CIIC 90/2020 of Universidad
de Guanajuato, Mexico. We acknowledge the work and support of the developers of the following following python packages: Astropy \citep{astropy:2013, astropy:2018} and Numpy \citep{vaderwalt_numpy_2011}. We also made extensive use of the visualisation and analysis packages DS9\footnote{\href{ds9.si.edu}{http://ds9.si.edu/site/Home.html}} and Topcat \citep{2005ASPC..347...29T}. We thank an anonymous referee for their insightful comments that have improved the manuscript. This work was compiled in the very useful free online \LaTeX{} editor Overleaf.

\bibliographystyle{pasa-mnras}
\bibliography{bib.bib}

\appendix
\label{sec:appendix}
\begin{table*}
    \centering
    \caption{Column descriptions corresponding to the supplementary electronic table. The MIDAS\_Name is derived using the MIDAS\_RA and MIDAS\_Dec columns. Entries without a GAMA\_IAUID, mag\_Ks, or NED\_Object\_Name are assigned a value of '$-99$'. Hotspots (L1) and Hotspots (L2) refer to the number of GLASS hotspots found in each lobe, where L1 and L2 refer to the lobe and counter-lobe. We define L1 as having the smaller position angle measured in a counter-clockwise direction from the North.}
    \label{tab:supplementarytable}
    \begin{tabular}{lcccccc}
    \hline 
    
    \hline
    No. & Column name & Unit & Description \\
    \hline
    \hline
    1 & MIDAS\_name & J~hh:mm:ss-dd:mm:ss & Name of radio source in J2000 format. \\
    2 & MIDAS\_RA & deg & Right Ascension of MIDAS source. \\
    3 & MIDAS\_Dec & deg & Declination of MIDAS source. \\
    4 & AGN\_status & -- & Activity state of AGN.\\
    5 & FR\_classification & -- & Fanaroff \& Riley Classification: FR-I, FR-II. \\
    6 & LAS & arc-seconds & Largest angular size measured from EMU-ES. \\
    7 & Hotspots (L1) & -- & Number of GLASS hotspots present in L1. \\
    8 & Hotspots (L2) & -- & Number of GLASS hotspots present in L2. \\
    9 & peak\_flux\_core & $\mu$Jy/beam & 5.5\,GHz radio core peak flux density.\\
    10 & err\_peak\_flux\_core & $\mu$Jy/beam & Error on 5.5\,GHz radio core peak flux density.\\
    11 & S$_{118}$ & mJy & Integrated flux density at MHz.\\
    12 & err\_S$_{118}$ & mJy & Error on integrated flux density at MHz.\\
    13 & S$_{154}$ & mJy & Integrated flux density at MHz.\\
    14 & err\_S$_{154}$ & mJy & Error on integrated flux density at MHz.\\
    15 & S$_{186}$ & mJy & Integrated flux density at MHz.\\
    16 & err\_S$_{186}$ & mJy & Error on integrated flux density at MHz.\\
    17 & S$_{216}$ & mJy & Integrated flux density at MHz.\\
    18 & err\_S$_{216}$ & mJy & Error on integrated flux density at MHz.\\
    19 & S$_{399}$ & mJy & Integrated flux density at MHz.\\
    20 & err\_S$_{399}$ & mJy & Error on integrated flux density at MHz.\\
    21 & S$_{887}$ & mJy & Integrated flux density at MHz.\\
    22 & err\_S$_{887}$ & mJy & Error on integrated flux density at MHz.\\
    23 & S$_{1400}$ & mJy & Integrated flux density at MHz.\\
    24 & err\_S$_{1400}$ & mJy & Error on integrated flux density at MHz.\\
    25 & S$_{2100}$ & mJy & Integrated flux density at MHz.\\
    26 & err\_S$_{2100}$ & mJy & Error on integrated flux density at MHz.\\
    27 & S$_{5500}$ & mJy & Integrated flux density at MHz.\\
    28 & err\_S$_{5500}$ & mJy & Error on integrated flux density at MHz.\\
    29 & S$_{9000}$ & mJy & Integrated flux density at MHz.\\
    30 & err\_S$_{9000}$ & mJy & Error on integrated flux density at MHz.\\
    31 & CP & -- & 216\,MHz core prominence.\\
    32 & GAMA\_IAUID & -- & ID of host galaxy in GAMA photometry catalogue. \\
    33 & Host\_RA & deg & Right ascension of host galaxy.\\
    34 & Host\_Dec & deg & Declination of host galaxy.\\
    35 & z & -- & Redshift.\\
    36 & z\_type & --& Redshift type: Spectroscopic, Photometric, lower-limit. \\
    37 & mag\_K$_s$ & -- & Host galaxy VIKING K$_s$-band magnitude.\\
    38 & NED\_Object\_Name & -- & Name of NED object corresponding to the GAMA\_IAUID. \\
    39 & log10\_L$_{216}$ & log10(W\,Hz$^{-1}$) & 216\,MHz radio power.\\
    40 & LLS & kpc & Largest linear size.\\

     \hline
     
     \hline
    \end{tabular}
\end{table*}

\end{document}